\RequirePackage{fix-cm}
\documentclass[smallcondensed,natbib]{svjour3}     
\smartqed  
\usepackage{graphicx}
\newcommand{\kmprs}  {\mbox{\rm km\,s$^{-1}$}}
\newcommand{\feh} {\mbox{\rm [Fe/H]}}
\newcommand{\mh} {\mbox{\rm [M/H]}}

\newcommand{\xh} {\mbox{\rm [X/H]}}

\newcommand{\oh} {\mbox{\rm [O/H]}}

\newcommand{\mgh} {\mbox{\rm [Mg/H]}}

\newcommand{\co} {\mbox{\rm [C/O]}}
\newcommand{\conum} {\mbox{\rm C/O}}

\newcommand{\mgsinum} {\mbox{\rm Mg/Si}}
\newcommand{\xfe} {\mbox{\rm [X/Fe]}}
\newcommand{\xmg} {\mbox{\rm [X/Mg]}}

\newcommand{\cfe} {\mbox{\rm [C/Fe]}}
\newcommand{\cmg} {\mbox{\rm [C/Mg]}}
\newcommand{\cm} {\mbox{\rm [C/M]}}

\newcommand{\nm} {\mbox{\rm [N/M]}}
\newcommand{\ofe} {\mbox{\rm [O/Fe]}}
\newcommand{\omg} {\mbox{\rm [O/Mg]}}
\newcommand{\nafe} {\mbox{\rm [Na/Fe]}}
\newcommand{\namg} {\mbox{\rm [Na/Mg]}}
\newcommand{\mgfe} {\mbox{\rm [Mg/Fe]}}
\newcommand{\femg} {\mbox{\rm [Fe/Mg]}}

\newcommand{\alfe} {\mbox{\rm [Al/Fe]}}
\newcommand{\almg} {\mbox{\rm [Al/Mg]}}
\newcommand{\sife} {\mbox{\rm [Si/Fe]}}
\newcommand{\simg} {\mbox{\rm [Si/Mg]}}

\newcommand{\cafe} {\mbox{\rm [Ca/Fe]}}
\newcommand{\camg} {\mbox{\rm [Ca/Mg]}}

\newcommand{\tife} {\mbox{\rm [Ti/Fe]}}
\newcommand{\timg} {\mbox{\rm [Ti/Mg]}}
\newcommand{\crfe} {\mbox{\rm [Cr/Fe]}}
\newcommand{\crmg} {\mbox{\rm [Cr/Mg]}}
\newcommand{\mnfe} {\mbox{\rm [Mn/Fe]}}
\newcommand{\mnmg} {\mbox{\rm [Mn/Mg]}}
\newcommand{\nife} {\mbox{\rm [Ni/Fe]}}
\newcommand{\nimg} {\mbox{\rm [Ni/Mg]}}
\newcommand{\cufe} {\mbox{\rm [Cu/Fe]}}
\newcommand{\cumg} {\mbox{\rm [Cu/Mg]}}
\newcommand{\znfe} {\mbox{\rm [Zn/Fe]}}
\newcommand{\znmg} {\mbox{\rm [Zn/Mg]}}

\newcommand{\bafe} {\mbox{\rm [Ba/Fe]}}
\newcommand{\srfe} {\mbox{\rm [Sr/Fe]}}
\newcommand{\eufe} {\mbox{\rm [Eu/Fe]}}

\newcommand{\bay} {\mbox{\rm [Ba/Y]}}
\newcommand{\yba} {\mbox{\rm [Y/Ba]}}
\newcommand{\ymg} {\mbox{\rm [Y/Mg]}}

\newcommand{\bamg} {\mbox{\rm [Ba/Mg]}}

\newcommand{\alphafe} {\mbox{\rm [$\alpha$/Fe]}}
\newcommand{\teff}  {\mbox{$T_{\rm eff}$}}
\newcommand{\Tc}  {\mbox{$T_{\rm c}$}}

\newcommand{\logg}  {\mbox{{\rm log}\,$g$}}

\newcommand{\turb}  {\mbox{$\xi_{\rm turb}$}}

\newcommand{\LiI} {Li{\sc\,i}}

\newcommand{\OI} {O{\sc\,i}}
\newcommand{\oI} {\mbox{\rm [{O{\sc\,i}}]}}
\newcommand{\MgI} {Mg{\sc\,i}}
\newcommand{\NaI} {Na{\sc\,i}}
\newcommand{\AlI} {Al{\sc\,i}}

\newcommand{\CaI} {Ca{\sc\,i}}
\newcommand{\CaII} {Ca{\sc\,ii}}

\newcommand{\TiII} {Ti{\sc\,ii}}
\newcommand{\CrI} {Cr{\sc\,i}}
\newcommand{\CrII} {Cr{\sc\,ii}}

\newcommand{\FeI} {Fe{\sc\,i}}
\newcommand{\FeII} {Fe{\sc\,ii}}

\newcommand{\CuI} {Cu{\sc\,i}}

\newcommand{\BaII} {Ba{\sc\,ii}}
\newcommand{\ThII} {Th{\sc\,ii}}
\newcommand{\NdII} {Nd{\sc\,ii}}

\newcommand{\Vtotal}   {\mbox{$V_{\rm total}$}}

\newcommand{\lt} {\mbox{$\, \langle \,$}}
\newcommand{\gt} {\mbox{$\, \rangle \,$}}
\def\ltsima{$\; \buildrel < \over \sim \;$}
\def\simlt{\lower.5ex\hbox{\ltsima}}
\def\gtsima{$\; \buildrel > \over \sim \;$}
\def\simgt{\lower.5ex\hbox{\gtsima}}
\journalname{Astron Astrophys Rev}
\begin{document}

\title{High-precision stellar abundances of the elements 
}
\subtitle{Methods and applications}


\author{Poul Erik Nissen         \and
        Bengt Gustafsson
}


\institute{P. E. Nissen \at
Stellar Astrophysics Centre, Department of Physics and Astronomy,
Aarhus University, Ny Munkegade 120, 8000 Aarhus C, Denmark. \\     
              \email{pen@phys.au.dk}           
           \and
           B. Gustafsson \at
Department of Physics and Astronomy, Uppsala University, Box515,
751\,20 Uppsala and Nordita, Stockholm, Sweden. \\
              \email{Bengt.Gustafsson@physics.uu.se}
}

\date{Received: date / Accepted: date}

\maketitle

\begin{abstract}
Efficient spectrographs at large telescopes have made
it possible to obtain high-resolution spectra of stars with high signal-to-noise
ratio and advances in model atmosphere analyses have enabled estimates
of high-precision differential abundances of the elements from these spectra,
i.e. with errors in the range 0.01 - 0.03\,dex for F, G, and K stars. 

Methods to determine such high-precision abundances together with precise values of 
effective temperatures and surface gravities from equivalent widths of spectral lines or 
by spectrum synthesis techniques are outlined, and
effects on abundance determinations from using a 3D non-LTE 
analysis instead of a classical 1D LTE analysis are considered. 

The determination of high-precision stellar abundances
of the elements have led to the discovery of unexpected phenomena
and relations with important bearings on the astrophysics of galaxies, stars,
and planets, i.e.
$i)$ Existence of discrete stellar populations within 
each of the main Galactic components (disk, halo, and bulge) providing new constraints
on models for the formation of the Milky Way. 
$ii)$ Differences in the relation between
abundances and elemental condensation temperature for the Sun and solar twins
suggesting dust-cleansing effects in proto-planetary disks and/or engulfment
of planets by stars;  
$iii)$ Differences in chemical composition between
binary star components and between members of open or globular clusters
showing that star- and cluster-formation processes are more complicated than previously thought; 
$iv)$ Tight relations between some abundance ratios and age for solar-like stars providing
new constraints on nucleosynthesis and Galactic chemical evolution models as well as
the composition of terrestrial exoplanets.

We conclude that if stellar abundances with precisions of 0.01 - 0.03\,dex can be
achieved in studies of more distant stars and stars on the giant and supergiant
branches, many more interesting future applications, of great relevance to stellar and galaxy
evolution, are probable. Hence, in planning abundance surveys, it 
is important to carefully balance the need for large samples of stars 
against the spectral resolution and signal-to-noise ratio needed to obtain high-precision
abundances. Furthermore, it is an advantage to work 
differentially on stars with similar atmospheric parameters,
because then a simple 1D LTE analysis of stellar spectra may be sufficient. However, when
determining high-precision absolute abundances or differential abundance between stars having
more widely different parameters, e.g. metal-poor stars compared to the Sun or giants to dwarfs,
then 3D non-LTE effects must be taken into account.

\end{abstract}

\section{Introduction}
\label{intro}
The chemical analysis of stars by stellar spectroscopy provides basic data of key significance 
for a broad range of areas in contemporary astronomy and astrophysics, extending from the study
of stellar evolution, of the formation of planetary systems, of Galactic and galaxy evolution, 
to cosmology. The basis of the analysis was founded already in the early 19th century on the
discovery of the dark Fraunhofer lines in the solar spectrum, the line identifications with
terrestrial elements some decades later, and the first observations of stellar spectra in
the second half of the century. However, it was not until the progress in atomic and molecular
physics during the first half of the 20th century that quantitative chemical analysis was
possible and gave important results. In recent decades, the methods of analysis have been
developed impressively, due to progress in three areas: One is the advent of telescopes
with large light collecting area, comparatively effective spectrometers and sensitive
detectors with linear response to illumination. Earlier, the low efficiency of instruments
and the non-linear properties of photographic emulsions plagued the analyses --  now it is
possible to obtain spectra for hundreds of thousands of stars of a quality which 50 years ago
were only available for the Sun and a handful of bright stars. Very important
are also the advances in laboratory measurements and calculations of accurate atomic and
molecular data, in particular line identifications, oscillator strengths and other relevant
cross sections. Previous data, e.g. for oscillator strengths for lines of the most important
elements, could suffer from systematic errors of a full order of magnitude, if available at all.
The progress in the third respect, the theoretical modelling of the spectrum-emitting regions,
the stellar atmospheres, and of their predicted spectra, is as impressive. It has now become
possible to construct models that are as physically consistent as can be reasonably required:
to consider the effects of the spectral lines on the structure of the atmospheres in full detail,
to replace the previous assumptions of plane-parallel stratification and simplified models of
the convective energy transport (primarily the Mixing-Length ``Theory", MLT) with a full
solution in 3D of the hydrodynamic equations as well as the radiative energy transfer,
and in the spectrum calculation to adopt the realistic velocity fields from the hydrodynamic
simulations and also abandon the previously made Local Thermodynamic-Equilibrium (LTE) assumption
-- that is to allow for all the non-local radiative effects on the excitation and de-excitation
of each atomic level, and on the ionisation and recombination of each ionisation stage,
of significance (in the literature denoted ``non-LTE''). 
In practice, these theoretical advances have led to
improvements which in many cases are decisive, not the least for the analysis of metal-poor
stars where the thermal inhomogeneities and the non-local radiative fields may strongly affect
the visible stellar spectrum. These improvements are well described by \citet{Asplund05}.

Since long ago, the errors in spectroscopic stellar-abundance determinations have been 
disputed, but often systematically underestimated. The presence of systematic errors in 
observations, interpretations and models have been hard to take into account fully, since there has 
been an absence of reliable comparisons for error calibration; standard spectra and analyses have 
also suffered from error sources similar to those of the programme stars like saturation effects, 
misidentified or unidentified spectral lines, systematic errors in physical
line-data, and severe departures from the model-assumptions of LTE and MLT. In this situation, the 
error estimates in standard abundance analyses have most often been intrinsic, based on the scatter 
in results for different individual spectral lines, at best also representing different species 
(ions, molecules) in which the element in question is present. Other sources of error which are 
also usually considered in these estimates are the errors in fundamental parameters such as 
effective temperature \teff\ and surface gravity $g = G\,M / R^2$ 
(where $M$ is stellar mass, $R$ radius and $G$ 
is Newton's constant of gravity). Sometimes also the estimated uncertainties in fudge parameters, 
in particular the microturbulence parameter \turb\ that is needed when partially saturated  
spectral lines are used in standard 1D model analyses, are included in the error analyses. 
Also, in the more refined 3D analyses in non-LTE, a number of errors in physically well-defined 
quantities such as collision cross sections and dissociation
energies may be important and have to be included in the discussion of abundance errors.

Just as the spectral abundance analyses have improved considerably during recent years, one must 
expect that the error analyses have improved, not the least due to the considerably more consistent 
models as such, which, even if they are yet not in regular use, serve as comparison when judging 
what errors are to be expected in standard analyses. During the last decade, a number of analyses 
have been published where such a detailed discussion has demonstrated errors smaller than 0.04\,dex 
in the logarithm of the abundance. In all 
cases, these are errors in relative abundances, i.e. as compared with a standard star, 
such as the Sun 
(and relative to hydrogen as the dominating element). The absolute determination of abundances is 
more difficult. In the global study of solar abundances by \citet{asplund09} typical abundance 
errors range from 0.04 to 0.10\,dex.
The reason why the relative errors in stellar abundances
may be smaller than those solar absolute ones is that some 
systematic errors, notably those related to oscillator strengths and
model uncertainties, may be assumed to cancel 
efficiently if the star is physically not too different from the standard star. This approach is 
basically built on the finding that stellar spectra may indeed be characterized by a small number 
of stellar fundamental parameters. The degree to which it may be applied will be
further discussed below.

Before proceeding into a more technical discussion of the reliability of claimed high-precision 
abundance determinations, it is of interest to consider what use such high precisions
(errors less than 0.04\,dex) could have. 
When summarizing the discussion of an ESO stellar-abundance workshop in 1980, celebrating the new 
ESO CAT/CES spectrometer with its linear Reticon detector arrays which brought us several important 
steps forward in these analyses, Bernard Pagel made the interesting distinction between two 
categories of stellar spectroscopists: the ``broad sweepers'' and the ``ultimate refiners''. The broad 
sweepers were primarily interested in global Galactic evolution problems, and were indeed often 
satisfied with an accuracy on the order of 0.2\,dex in their abundance results, while the ultimate 
refiners wanted more accurate data, often driven by stellar physics considerations. This situation 
is now changed. A number of discoveries and ideas have increased the interest in high-precision 
abundances, even among scientists that are not primarily interested in stars in themselves. One 
could, in principle, see this interest as a result of an often-made experience in natural sciences: 
when the accuracy of measurements is increased significantly, new important structures in nature are 
disclosed. Among these proposed or discovered new structures are the following:
\begin{enumerate}
\item The possibility to distinguish different stellar components, or sub-populations, in the Galaxy 
by finding certain common special abundance profiles among stars 
\citep[``chemical tagging'',][]{freeman02} 
and from that infer and map the early history of the Galaxy, with in-falling 
and merging minor systems. This idea has got a wider application, also for studies of the evolution 
of the Galactic thin and thick disks and its bulge with some interesting results. 
The field is further stimulated 
by systematic combination of data from the astrometric satellite {\em Gaia} with high-resolution 
spectroscopy.
\item The systematic comparison of solar twins\,\footnote{A
solar twin is usually defined as a star having atmospheric parameters agreeing with
those of the Sun within $\pm 100$\,K in effective temperature and $\pm 0.10$\,dex in surface gravity and
metallicity.} with the Sun which has unexpectedly  proven that the 
Sun is somewhat poor in refractory elements as compared with most twins in the solar neighbourhood 
\citep{melendez09}. This finding, where the Sun departs in composition as a function of the 
dust-condensation temperature of the element in question, has opened up the possibility to study 
the conditions in the environment where the Sun was once formed, and a search for “Solar siblings” 
in the Galaxy.
\item The study of the possibly different chemical composition of the components of binaries.  
\citet{laws01} and \citet{ramirez11} found that the secondary star of 16 Cyg, 
known to host a giant planet, is slightly 
metal-poor relative to the primary, and similar effects have later been found for other binaries
\citep[e.g.][]{saffe17, oh18}. It 
seems that these abundance differences give information on the early evolution of these systems, 
including their planets.
\item The study of different chemical composition among stars in clusters.  
For long, differences, including anti-correlations, between the abundances of the 
light elements, like C, O, Na, Mg and Al have been noticed for globular clusters 
\citep{cohen78, peterson80, kraft93, gratton01b}, and interpreted in terms of
self-enrichment by matter from early generations of stars in the forming cluster.
Several details
in such a scenario are, however, not understood and depend on further detailed studies. Similarly,
the relatively small effects of differential diffusion of heavy elements in the stars were traced
by \citet{korn07} for the cluster NGC 6397 and such studies have been pursued to other more metal-rich
clusters \citep[e.g.][]{onehag14}. For open clusters several studies
have been devoted to the search for abundance differences among cluster stars with negative or 
non-significant results. A first sign of chemical inhomogeneities at the 0.02 - 0.04\,dex level among the 
solar-type stars in the Hyades and Pleiades clusters was, however, reported by \citet{liu16b}
and \citet{lorenzospina18}.
\item Chemical abundances for the study of nucleosynthesis and stellar evolution. In general, the 
study of the composition of stars of different ages and at different location in the Galaxy (and 
other galaxies) lead to information on mixing processes in the interstellar medium
as well as in the stars when nuclear reactions are brought to the stellar surface.
The very accurate measurement of these abundances are sometimes of key
significance for the understanding of the underlying processes, see e.g.
review by \citet{nomoto13}.
\item The search for abundance signatures in planetary hosts of different kinds. Early on it was
realized that stars of solar type with known massive planets were comparatively metal-rich
\citep{gonzalez01, santos01}
but a similar effect for different types of stars and planetary systems is
disputed. Other effects of planets, on the abundances of lithium and other elements
in the planetary hosts are studied with ambiguous results. Conversely, abundances in
stars set constraints on the composition of proto-planetary disks and hence on
the composition of planets.

\end{enumerate}

These examples of discoveries when high-precision abundance analyses are applied 
also suggest new possibilities for future important work on other types of stars or on more 
distant stars for which very accurate analysis may be attainable within a near future. 
However, to achieve
an accuracy of 0.03\,dex or better, improved understanding of the structure and
dynamics of the atmospheres of the corresponding stars will be needed 
together with sufficiently complete atomic and 
molecular data, as well as adequate observing time at larger telescopes 
in order to secure sufficiently high S/N and spectral resolution. Then,  the 
chemical composition of red giant stars, AGB stars and even supergiants
may be explored in detail and used in the study of more distant regions, in the Galaxy and
in external galaxies.

\section{Methods}
\label{methods}
As a first step towards high-precision abundances of elements in stars, 
one has to obtain stellar spectra with high resolution and high 
signal-to-noise. Spectral lines  
are Doppler broadened due to thermal and
turbulent motions of absorbing atoms in stellar atmospheres,
which for late-type stars corresponds to a profile with a FWHM of $\sim 6$\,\kmprs . 
Hence, in order to avoid a significant additional instrumental broadening of spectral lines,
one needs a spectrograph with a
resolution, $R = \lambda/\Delta\lambda _{\rm instr} \simgt 50\,000$, 
where $\Delta\lambda _{\rm instr}$
is the FWHM of the instrumental profile. This makes it easier to
disentangle blended lines and define a reliable 
continuum. However, lower resolution
may also be used to determine high-precision 
abundances when using spectrum synthesis techniques, if an accurate and
sufficiently complete list of lines
is available for the spectral region(s) studied.

\subsection{Equivalent width measurements and abundance errors}
\label{EW}
In most high-precision studies, abundances have been derived from equivalent widths (EWs)
measured by fitting Gaussian profiles to the spectral lines or by direct integration, i.e.
\begin{eqnarray}
EW \, = \, \sum_{i=1}^{n_{pix}} \frac{F_c - F(\lambda_i)}{F_c} \, \delta \lambda_{pix},
\end{eqnarray}
where $F_c$ is the continuum flux, $F(\lambda_i)$ the flux in the line
at a particular wavelength point $\lambda_i$, $\delta \lambda_{pix}$
the spectral pixel size, and $n_{pix}$ the number of pixels aross the line.
For a given value of $F_c$, the statistical error of the EW is then
\begin{eqnarray}
\sigma (EW) \, \simeq \, \sqrt{n_{pix}} \, (S/N)^{-1} \, \delta \lambda_{pix},
\end{eqnarray}
where $S/N$ is the signal-to-noise ratio per spectral pixel assumed to be
approximately constant across the line.

With three pixels per spectral resolution element 
($\Delta\lambda _{\rm instr} = 3\,\delta \lambda_{pix}$) 
as in many high-resolution echelle spectrographs,
we get $\delta \lambda_{pix} = 0.04$\,\AA\ at $\lambda = 6000$\,\AA\ for a
resolution of $R = 50\,000$. The flux in the line is typically integrated over
a width of 0.4\,\AA\ corresponding to $n_{pix} \simeq 10$. Hence, a S/N of 200 is
sufficient to measure the EW with a statistical error of 0.6\,m\AA , 
which for a line with $EW \simeq 25$\,m\AA\ corresponds to a relative error of 2.4\,\%.
As the abundance derived from such a weak line (on the linear part of the
curve of growth) is nearly proportional to the equivalent width, the
corresponding error in logarithmic abundance is $\sim 0.01$\,dex.

Uncertainties in the continuum setting may introduce larger errors.
A shift of the continuum by 0.5\,\% changes the EW
by 2\,m\AA\ in the example given above, i.e. more than three times the statistical
error given by Eq. (2). The continuum effect on differential abundances can, however,
be kept below 0.01\,dex, if EWs are measured relative to the same continuum windows 
and a set of spectral lines having similar strengths for 
all programme stars are used.
If, on the other hand, a wide range in abundances is to be covered, large differences
in line strength are present and it may not
be possible to use the same set of spectral lines in all stars, which means that 
the abundances derived are sensitive to the continuum setting and are
also affected by errors in the $gf$-values. For this reason,
trends of abundance ratios as a function of  \feh\ are not as precise as
differential abundances at a given metallicity.

As shown by \citet{bedell14}, it is important to observe all program stars 
with the same spectrograph at a fixed resolution, if differential abundances
with a precision of $\sim 0.01$\,dex are to be obtained. This requirement raises special problems
if the Sun is used as a standard, as is often the case, because the solar flux spectrum cannot 
be observed in direct sunlight in the same way as the stellar spectra. 
One may, however, use reflected sunlight from asteroids
to obtain a good proxy of the solar flux spectrum.
\cite{bedell14} show that abundances derived from $S/N \sim 700$ spectra 
of Vesta, Iris, and Ceres observed
with the same spectrograph agree with a rms scatter of 0.006\,dex. A similar good
agreement between spectra of
Vesta, Victoria and the Jupiter moon Europa observed with the HERMES spectrograph 
at the 1.2\,m Mercator telescope was obtained by \citet{beck16}.
Furthermore, they found
that the normalized spectrum of Europa ($S/N \sim 450$) agrees with the revised Kitt Peak Solar Flux
Atlas of \citet{kurucz05} (degraded to the resolution of HERMES, $R = 85\,000$) with a rms 
deviation of 0.25\,\% over the spectral range 4000 - 8000\,\AA\ (excluding regions with
strong telluric lines). This suggests that such reflected sunlight spectra are adequate
for determining differential stellar abundances with respect to the Sun at a precision
better than 0.01\,dex, but further investigations should be made, given that the Sun is found
to have an unusual ratio between volatile and refractory elements  
compared to solar twins as discussed in Sect. \ref{solartwins}.

\begin{figure}
  \includegraphics[width=12.4cm]{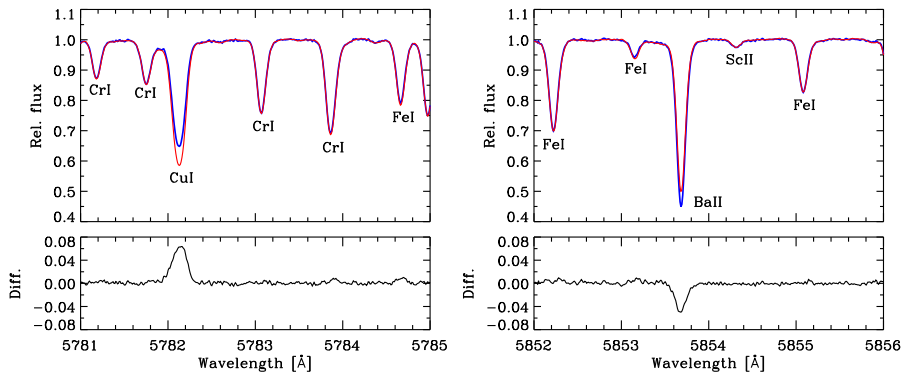}
\caption{Solar flux ESO 3.6\,m HARPS spectrum
(red line) in comparison with the HARPS spectrum of HD\,96116 (blue line)
near the \CuI\ line at 5782.12\,\AA\ (left panel) and the \BaII\ line at 5853.70\,\AA\
(right panel).
The lower panels show the difference (HD\,96116 -- Sun) between the two spectra.
Based on spectra from \citet{nissen16}.}
\label{fig:5782-5853}
\end{figure}

As an illustration of the differential way of measuring equivalent widths, 
Fig.\,\ref{fig:5782-5853} shows the ESO 3.6\,m HARPS spectrum ($R = 115\,000, S/N \simeq 600$)
of the young solar twin star, HD\,96116, in comparison with the solar flux HARPS spectrum
($S/N \sim 1200$) observed in reflected sunlight from the asteroid Vesta.
As seen the \CrI\ and \FeI\ lines have nearly the same strengths in the two stars,
whereas the \BaII\ line is stronger for HD\,96116 ($\Delta EW = 7.7$\,m\AA )
and the \CuI\ line is weaker ($\Delta EW = -13.7$\,m\AA ).  
Taking into account the small differences in atmospheric parameters 
between HD\,96116 and the Sun, \citet{nissen16} has used 
the differences in EWs to derive \bafe \,\footnote{For two elements, A and B,
with number densities $N_{\rm A}$ and $N_{\rm B}$, 
[A/B] $\equiv {\rm log}(N_{\rm A}/N_{\rm B})_{\rm star}\,\, - 
\,\,{\rm log}(N_{\rm A}/N_{\rm B})_{\rm Sun}$.} = 0.190 and
$\cufe = -0.098$ for HD\,96116 in a study of
trends of abundance ratios as a function of stellar age.

In most high-precision abundance studies \citep[e.g.][]{melendez09,
ramirez14, tuccimaia14, nissen15, spina16b}, equivalent widths have been
measured by interactive fitting of weak to medium-strong lines
($EW \simlt 80$\,m\AA\  in the visual spectral region)  by Gaussian profiles
using the IRAF {\tt splot} task.
This has the advantage that problems of line blending can be directly seen
and eventually solved by using the deblending routine in {\tt splot}.
For very weak lines and lines with asymmetric or box-like
profiles due to hyper-fine-structure (HFS), direct integration of the line is
to be preferred. Strong lines with damping wings should be
measured by Voigt profile fitting, but the resulting EWs are sensitive to weak
blends in the wings and to the continuum setting, so such lines should be avoided
if possible.

Interactive measurements of EWs are, however, time-consuming and
unsuitable for large surveys in which millions of lines have to be 
measured\,\footnote{In an abundance study of 714 Galactic disk
stars \citep{bensby14}, a record high number of 300\,000 spectral lines were measured
interactively by Thomas Bensby using the IRAF {\tt splot} task}. 
Programs for automated measurement of EWs have therefore been developed, e.g.
DAOSPEC \citep{stetson08}, iSpec \citep{blanco14}, and ARES v2 \citep{sousa15}.
and extensively applied for stellar abundance studies. It would be
interesting to see how well high precision abundances based on EWs measured by such programs
agree with abundances based on interactive measurements of EWs.

\subsection{Stellar parameters}
\label{param}
In order to derive high-precision stellar abundances, the effective temperature, \teff ,
and surface gravity, $g$, of a star should also be known with high precision. For late-type
stars, abundances derived from equivalent widths of low-excitation neutral metal-lines
change by up to 0.1\,dex if \teff\ of the model atmosphere used to analyse the lines
is changed by 100\,K, and abundances derived from ionized metal-lines may be subject
to a similar change if \logg\ is changed by 0.3\,dex. Hence, the stellar parameters
should be determined with precisions of $\sigma (\teff) \, \simeq 10$\,K and
$\sigma (\logg) \, \simeq  0.03$\,dex to reach precisions of $\sim 0.01$\,dex in
the abundance determinations. Such high precisions
may be obtained by analysing EWs of the many Fe lines
available in late-type stellar spectra. The \teff\ of the model
atmosphere is often chosen so that there
is no systematic dependence of the derived Fe abundance on excitation potential, 
$\chi_{\rm exc}$, of the lines,
and \logg\ may be determined from the requirement that the Fe abundances derived from 
\FeI\ and \FeII\ lines should agree. As the electron pressure in the models is 
significantly affected by the abundance of Fe and some 
alpha-capture elements (especially Mg and Si), one must also assure that the models applied
have consistent values of \feh\ and \alphafe . This means that the  determination of \teff\ and 
\logg\ is an iterative process, where the final parameters are found by interpolation in 
a 4-dimensional grid of model atmospheres.

Effective temperatures can also be determined by the infrared flux method (IRFM)
or from colour indices like $V-K$ as discussed in detail by \citet{casagrande10}.
For groups of nearby stars with similar parameters the precision  
is $\sim 30$\,K, but interstellar reddening is a problem for
more distant stars. Alternatively, one may obtain reddening-free \teff -values
from the wings of Balmer lines by comparing with model atmosphere profiles calculated 
from the theory of Stark broadening by charged particles and self-broadening 
by hydrogen atoms \citep[see review by][]{Barklem16R}. As shown by \citet{cayrel11},
\teff\ values derived from H$\alpha$ are offset by $\sim -100$\,K
relative to interferometric temperatures based on angular diameter and bolometric flux
measurements of solar-type stars (including the Sun). 
This may be due to problems with the
temperature structure of the 1D model atmospheres applied for the analysis of the
H$\alpha$ profiles, as is demonstrated by the fact that
the results obtained depend critically on the assumed mixing-length 
parameter of the 1D models \citep{ludwig09a}.  Recent 3D non-LTE modelling 
by \citet{amarsi18} leads to better agreement between Balmer-line temperatures
and interferometric values of \teff , and 
it is encouraging that the scatter in the relation between
interferometric and H$\alpha$ temperatures  after the 
offset has been applied is only 30\,K for stars with
$5000 < \teff < 6400$\,K and $-0.7 < \feh < 0.2$ \citep{cayrel11}.  
A similar scatter was found by \citet{ramirez14} when comparing H$\alpha$ temperatures
of solar twins stars with precise excitation temperatures.
Further improvements of the precisions of Balmer line temperatures to say
$\sigma (\teff ) \sim 10$\,K may, however, be difficult because the continuum of the observed
spectra have to be defined very precisely over regions of $\pm 30$\,\AA , which
is a considerable fraction of the free spectral range of an echelle-order in
a typical spectrograph. For this purpose, low-order grating spectrometers should be applied.

\begin{figure}
\centering
\includegraphics[width=9cm]{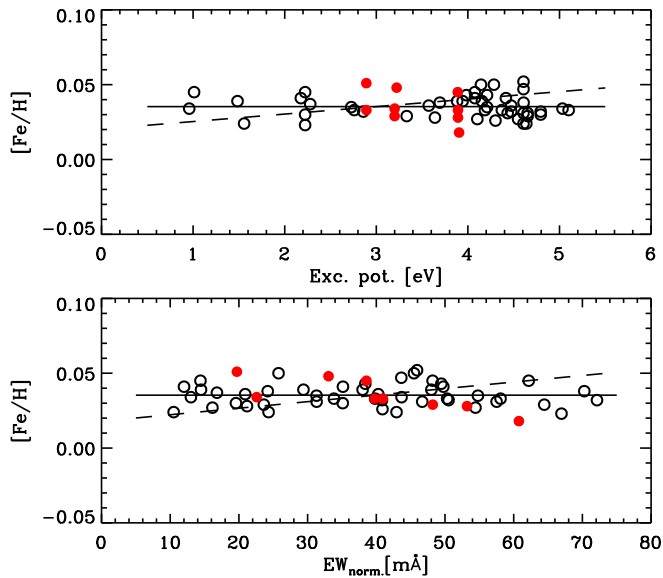}
\caption{[Fe/H] derived from EWs in the spectrum of the solar twin HD\,183658 as
a function of excitation potential (upper panel) and
equivalent width normalized to $\lambda = 6000$\,\AA ,
i.e. $EW_{\rm norm.} = EW / \lambda \cdot 6000.0$  (lower panel).
Open circles refer to \FeI\ lines and filled red circles to \FeII\ lines.
The full drawn lines show linear fits to the data for the derived
parameters of the star (\teff = 5809\,K, \logg = 4.402, \feh = 0.035, \turb = 1.04\,\kmprs ).
The dashed lines show changes in the slopes,
if \teff\ is decreased by 30\,K (upper panel) and the microturbulence
is decreased by 0.1\,\kmprs\ (lower panel). Based on data from \citet{nissen15}.}
\label{fig:183658}
\end{figure}

If  hydrostatic plane-parallel 1D model atmospheres are used, the effect of
gas motions on the equivalent widths has to be described by yet another parameter,
the microturbulence \turb , normally assumed to be independent of 
depth in the atmosphere and added in quadrature to the thermal Doppler velocity broadening.
The effect of this parameter on abundance determinations is negligible for weak lines,
say $EW \simlt 30$\,m\AA\ in the visual part of the spectrum, but becomes important for lines with 
$EW \simgt 50$\,m\AA . Hence,  \turb\ may be determined from the requirement
that the derived Fe abundances should be independent of EW. To avoid degeneracy between
the determination of \teff\ and \turb\ from equivalent widths, 
it is important that Fe lines in a small interval
of $\chi_{\rm exc}$ have a large range in EW. An example of the technique is shown
in Fig.\,\ref{fig:183658}.
 
From the 1-$\sigma$ errors of the slopes of \feh\ versus
$\chi_{\rm exc}$ and EW and the errors of the average Fe abundances derived from
\FeI\ and \FeII\ lines,  one can estimate the uncertainty of the atmospheric parameters
using error analysis methods, which take into account correlations in the 
parameter determinations \citep[e.g.][]{mcwilliam95, epstein10, bensby14}.
When based on equivalent width measurements in spectra with 
$R \simgt 50\,000$ and $S/N \simgt 400$, precisions in relative values
better than 10\,K in \teff , 0.02\,dex in \logg , and 0.01\,dex in \feh\ and \alphafe\
are estimated \citep[e.g.][]{melendez09, ramirez14, nissen15}. Comparisons of independent
determinations for solar twins confirm these estimates, but it should 
be emphasized that these small errors are only
relevant when stars with similar parameters are compared,
i.e. stars having ranges less than 300\,K in \teff\ and less than 
0.3\,dex in \logg\ and \feh . If larger differences are present, systematic errors due to
non-LTE and 3D effects may dominate and have to be considered,
see Sect. \ref{analysis}.

An alternative way of determining stellar surface gravities is based on
the fundamental relations between gravity, mass, radius, luminosity, and effective
temperature of a star, $g \propto {\cal M} / R^2$ and $R^2 \propto L / \teff ^4$,
corresponding to the equation
\begin{eqnarray} 
\log \frac{g}{g_{\odot}}  =  \log \frac{{\cal M}}{{\cal M}_{\odot}} +
4 \log \frac{\teff}{T_{\rm eff,\odot}} + 0.4 \, (M_{\rm bol} - M_{{\rm bol},\odot}),
\end{eqnarray}
where the various parameters are normalized on the solar values. 
The absolute bolometric magnitude, $M_{\rm bol}$, can be determined
from the apparent magnitude if the distance is known and the bolometric correction
can be estimated.  For a parallax error of 1\%,
the induced error of \logg\ will be about 0.01\,dex. With the Hipparcos
satellite such small parallax errors have been obtained only for a small fraction of
the solar twins,
binaries, and stars in clusters, for which high-precision abundances and
effective temperatures have been derived. Based on 
{\em Gaia} data, however, the parallaxes of most of the stars studied are expected to become known 
with accuracies better than 1\%. The largest contribution to the error
of \logg\ will then arise from the mass, which is often obtained by interpolating between 
stellar evolutionary tracks in the $L - \teff$ diagram. Knowing \teff\ and \feh\ from
high-precision spectroscopy, differential masses of solar twins can be obtained to
a precision of $\sim 0.01 \, {\cal M}_{\odot}$ \citep[e.g.][]{ramirez14}
corresponding to an error of 0.005\,dex in
\logg , but in other regions of the $L - \teff$ diagram the error may be much larger.

Precise surface gravities may also be determined from asteroseismology, i.e.
by comparing observed oscillation frequencies with  predictions from stellar models. 
For about 100 main-sequence and subgiant stars having long time series of
precise flux measurements obtained with the {\em Kepler} satellite, 
\logg\ has been determined with a precision better than 0.01\,dex 
\citep{silva-aguirre15, silva-aguirre17} and for K-giants in the
so-called APOKASC programme, \logg\ is determined with an average precision of 
0.014\,dex \citep{pinsonneault14}.

In addition to the methods discussed above two more spectroscopic methods 
for determining stellar gravities may be mentioned: the use of pressure-sensitive
wings of strong lines \citep{blackwell77} and the use of pressure-sensitive 
molecular lines like MgH \citep{bell85}.

The different ways of determining surface gravities make it possible to
investigate if variations in other model-atmosphere parameters than \teff , \logg , 
\feh , and \alphafe\ play a role when  determining precise abundances.
One such parameter is the helium abundance, $y = N_{\rm He} / N_{\rm H}$, which 
for example in the MARCS models \citep{gustafsson08} is assumed to be $y = 0.085$
corresponding to a helium mass fraction of $Y = 0.25$ 
close to the solar surface value determined from helioseismology \citep[e.g.][]{basu04}.
Given that atomic diffusion decreases the atmospheric helium abundance
as a function of increasing stellar age \citep[the effect is about 10\% for the Sun, e.g.][]{jcd93}
and Galactic chemical evolution increases the initial He abundance as a function of time, 
it would not be a surprise if young stars have $\Delta Y \sim 0.05$\,-\,0.10 
relative to the oldest stars. Such differences of the He abundance may
lead to a significant effect on spectral lines. For late-type stars, helium does not contribute to
opacity and free electrons, but it affects the mean molecular weight
and hence the electron pressure in the atmosphere of which, for instance,
the strengths of \FeII\ lines depend. 
As shown by \citet{stromgren82}, two models with identical temperature structures 
but different surface gravities, $g_1$ and $g_2$, and different helium abundances,
$y_1$ and $y_2$ (by number) will have closely similar $P_e (\tau )$ relations if
\begin{eqnarray} 
g_2 \, = \, g_1 \, \frac{1 + 4 y_1}{1 + y_1} \,\,\frac{1 + y_2}{1 + 4 y_2}.
\end{eqnarray}
This means that for two stars with the same \teff, same heavy element composition
relative to hydrogen,
and same equivalent widths of \FeII\ and \FeI\ lines, one will derive a
difference in gravity, $\Delta \logg \simeq -0.02$\,dex if the difference in
He mass fraction is $\Delta Y = 0.06$ (corresponding to an increase of the He
abundance from $y = 0.085$ to $y = 0.108$). Hence, a comparison of high-precision
spectroscopic gravities with gravities based on {\em Gaia} parallaxes or seismic data
may reveal possible variations in He abundances. 
In such a comparison, the effects of He abundance variations
on derived stellar masses (used for calculating trigonometric gravities)
and on seismic gravities should be taken into account.
The method may also be applied to determine He abundance differences
between stars in a cluster with the advantage that the distances do not
enter into the problem. Interestingly, \citet{lind11b} showed that 
the He abundance difference between a first and a second generation star
in the globular cluster NGC\,6397 is not very pronounced, i.e. $\Delta Y = 0.01 \pm 0.06$.

\subsection{Spectrum synthesis techniques}
\label{synthesis}
\begin{figure}
\centering
\includegraphics[width=10cm]{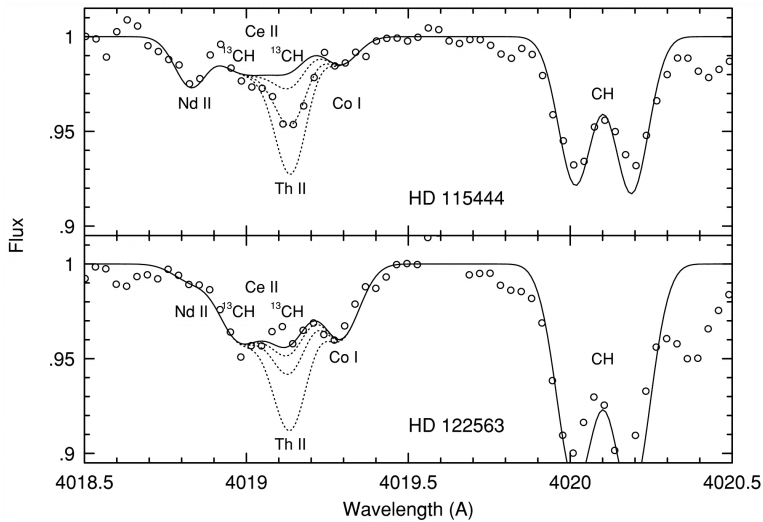}
\caption{A high-resolution ($R = 60\,000, S/N \simeq 200$) spectrum around the \ThII\ line at 4019.1\,\AA\
for the metal-poor $r$-process-rich K-giant star, HD\,115444 ($\feh \simeq -3.0$,
$\eufe = +0.9$) in comparison with the spectrum of a metal-poor,
$r$-process-poor K-giant star, HD\,122563 ($\feh \simeq -3.0$, $\eufe = -0.4$).
Full drawn lines show synthetic spectra without Th and
dotted lines with Th abundances of ${\rm log} \, \epsilon ({\rm Th}) = -2.48, -2.23, -1.98$ for
HD\,115444 and ${\rm log} \, \epsilon ({\rm Th}) = -3.08, -2.58, -2.08$ for HD\,122563.
Based on spectra from \citet{westin00}.}
\label{fig:westin}
\end{figure}

Instead of using equivalent widths, one may derive stellar abundances 
by comparing observed spectra with synthetic model-atmosphere spectra.
Pioneering studies were made by \citet{payne-gaposchkin40, baschek59,
fischel64, cayrel69, bell70}.
For heavily blended lines, this is the only way to determine precise abundances. 
An  example is provided in Fig.\,\ref{fig:westin}, which shows how the thorium abundance
of a very metal-poor $r$-process-rich K-giant star, HD\,11544, is determined by
\citet{westin00}. Although the \ThII\ line at 4019.14\,\AA\ overlaps with two
$^{13}$CH lines, the spectrum synthesis yields a fairly precise 
Th abundance of ${\rm log} \, \epsilon ({\rm Th}) = -2.23 \pm 0.05$\,\footnote{For an
element X, ${\rm log} \, \epsilon ({\rm X}) \equiv {\rm log} (N_{\rm X} / N_{\rm H}) + 12.0$},
whereas the \ThII\ line (and the \NdII\ line at 4018.84\,A) is not detected in the
comparison star HD\,122563.  As shown by Westin et al, this may be used to
estimate a radioactive age of HD\,115444 from the abundance of Th relative to
the abundance of stable $r$-process elements.

In the example 
shown in Fig.\,\ref{fig:westin}, the stellar atmospheric parameters were 
derived from equivalent widths by the method described in Sect. \ref{param}.
The stellar parameters may, however, also be determined by spectrum synthesis
techniques. A popular tool for this
is the Spectroscopy  Made Easy (SME) package \citep{valenti96, piskunov17}, which
can be used to interpolate in a grid of model atmospheres and to calculate
synthetic spectra based on a line list, e.g. the Vienna Atomic Line Database
(VALD\,\footnote{{\tt http://vald.astro.uu.se}. Note that in many cases it may be
proper and important to give a reference to the source of the original data and not only
refer to the data base.}). Atmospheric parameters
and abundances may then be derived by chi-square fitting of observed high-resolution spectra
in selected spectral windows. In addition to \teff , \logg , \feh , and microturbulence,
the fitting must include parameters for macroturbulent and rotational broadening
of spectral lines. These broadening mechanisms
do not change the total flux absorbed in a line and have therefore no direct effects
on parameters and abundances determined from equivalent widths, although there
may be indirect effects related to continuum placement and fitted line profiles or
integration limits when measuring the EWs.

An improvement of great significance in synthetic spectroscopy during recent
decades has been the precise quantum-mechanical calculation of collision broadening 
of atomic lines by \citet{anstee91, anstee95} and \citet{barklem00}
\citep[see][for further references]{Barklem16R}.

\citet{brewer16a} provide a good example of how stellar parameters and
elemental abundances can be determined with the SME package. Spectra of
$\sim 1600$ F, G and K main-sequence and subgiant stars were
obtained with the HIRES spectrograph at KECK Observatory with a resolution of 
$R \simeq 70\,000$ and $S/N \simgt 100$ for $\sim 75$\% of the sample. 
Synthetic spectra were calculated for specific wavelength regions between 5100 and 
7800\,\AA\ and fitted to the observed spectra. The strong \MgI \,b triplet in 
the 5150 - 5190\,\AA\ region was included, which
provides a measure of the surface gravity in addition to the information
obtained from the ratio of the strengths of \FeI\ and \FeII\ lines,
because the absorption in the wings  of the Mg lines increases with gas
pressure in the atmosphere. Atomic line data were adopted from VALD with
empirical adjustments of $gf$-values, van der Waals coefficients, and wavelengths
obtained by fitting the solar flux atlas of
\citet{wallace11} adopting abundances from \citet{grevesse07} and
a microturbulence of \turb = 0.85\,\kmprs .  Furthermore, the zero points
of [X/H] were checked and slightly adjusted using the average abundances derived from 20 spectra of four
asteroids observed with HIRES at the same resolution as the stars.
As a comment to this procedure, we note that
it is not easy to deduce absolute  $gf$-values or damping constants by fitting solar
spectra with accuracies better than 0.1\,dex in view of uncertainties in the solar model, 
departures from LTE, line blends, etc. In strictly differential analyses 
for solar-like stars, however, considerably higher accuracies may be achieved, 
since the effects of many of these sources of error will cancel. 

Another example of the use of the SME package is the analysis of
GALAH ($R = 28\,000$) spectra by \citet{buder18}, 
who determined atmospheric parameters and
abundances of 23 elements for a large sample of dwarf and
giant stars. For a number of key elements (Li, O, Na, Mg, Al, Si, and Fe) non-LTE corrections
were included, which makes the derived abundances more reliable.

Spectrum synthesis techniques, when applied to spectra containing partly saturated
lines, have the problem that the determination of the
microturbulence parameter is partially degenerate with 
the other parameters \teff , \logg , and \feh . When including \turb\ as as
free parameter, \citet{brewer15} find
deviations between seismic gravities and spectroscopic gravities of up to
0.25\,dex for a sample of 42 {\em Kepler} stars. Assuming instead a constant
microturbulence of \turb = 0.85\,\kmprs\ as adopted for the solar atmosphere,  
excellent agreement with the {\em Kepler} gravities is obtained, i.e. a rms deviation of only
0.05\,dex over a \teff\ range from 5000\,K to 6500\,K. This suggested constancy of
the microturbulence parameter is somewhat surprising, because from analyses
of EWs of main-sequence stars, \turb\ is found to increase by about 1.2\,\kmprs 
from \teff = 5000\,K to 6500\,K and depends also on gravity \citep[e.g.][]{edvardsson93, ramirez13}.
Another problem is that the average \feh\ abundance of stars 
in the sample of \citet{brewer16a} declines with about 0.15\,dex over the
\teff-range from 5800\,K to 5000\,K. This
may be related to the adoption of a constant
microturbulence and/or deficiencies in the stellar models and the 
assumption of LTE. Obviously, 3D non-LTE analyses are needed to get more
accurate abundances (see Sect. \ref{analysis}). 
However, this does not exclude that very precise differential
abundances can be derived for narrow ranges in \teff . From repeated observations
of stars having $S/N > 100$ spectra, \citet{brewer16a} obtain average std deviations of 
12\,K in \teff , 0.014\,dex in \logg , and $\simlt 0.01$\,dex for the abundance
of many elements.  

The APOGEE survey \citep{majewsky17} is another example of the use of spectrum 
synthesis techniques in deriving high-precision stellar abundances. During a 
three-year observing campaign on the Sloan 2.5\,m telescope, near-infrared spectra
with $S/N > 100$ covering the $H$-band (1.5 - 1.7\,$\mu$m) with a resolution of
$R \simeq 22\,500$ have been obtained for about 150\,000 stars. Most of them are
intrinsically bright K-giants and due to the relatively low interstellar 
absorption in the $H$-band, abundances in stars of various populations 
belonging to the Galactic disk, the bulge, and the halo can be studied over a wide
range of distances.

In the APOGEE survey, stellar parameters and abundances of up to 15 elements are 
derived by comparing observed
spectra to a large library of synthetic spectra using a chi-square minimization
programme ASPCAP \citep{garciaperez16}. The synthetic spectra are calculated for 
ATLAS9 model atmospheres \citep{Castelli03} and normalized to pseudo-continuum
spectral windows in the same way as the observed spectra. The line list \citep[see][]{shetrone15}
consists of more than 130\,000 atomic and molecular transitions for which the $gf$-values
were adjusted so that  high-resolution atlases of the Sun and 
Arcturus \citep{livingston91, hinkle95} are well fitted. 
Given that the $H$-band includes many molecular lines from CO, CN, and OH, 
parameters for the C, N, and $\alpha$-element abundances were included in the
grid of synthetic spectra in addition to the usual atmospheric parameters.
The microturbulence may be obtained from the fitting, 
but  a simple dependence on gravity,
as found for a subset of well observed stars, is adopted for K-giants. 
Instrumental broadening is applied to the calculated spectra taking into
account variations in the line spread function over the field of the CCD detectors.
Furthermore, a constant Gaussian broadening of 6\,\kmprs (FWHM) is adopted
to take care of macroturbulence. Rotational broadening is assumed to be negligible,
which may introduce errors in the derived parameters and abundances for stars
having $V {\rm sin} i \, \simgt \, 5$\,\kmprs , but this is only the case for a small
fraction of red giants.
 
After a first pass in which the atmospheric parameters of stars are determined from
the entire APOGEE spectral band, individual abundances of 15 elements  are
derived from selected spectral windows containing atomic and molecular lines of the
element considered. As discussed by \citet{holtzman15}, there are systematic deviations
in \logg\ relative to seismic gravities and trends in abundances as a function of \teff , 
but abundances derived for stars in solar metallicity clusters 
have only a small scatter in narrow ranges of effective temperature.
For the best observed elements (O, Mg, and Fe), the rms scatter in intervals of
$\Delta \teff = 250$\,K is about 0.03\,dex
for cluster stars having APOGEE spectra with $S/N \sim 200$ and 
for the abundance ratios \ofe\ and \mgfe , the scatter is only $\sim 0.02$\,dex
\citep{ness18}.

High-precision abundances may also be derived from low-resolution
($R \simlt 6000$) spectra, if accurate spectral models are available 
and accurate spectrum normalization can be achieved
\citep{ting17a}.
In crowded low-resolution spectra most absorption features and
pseudo-continuum windows are, however, affected by strong lines sensitive 
to microturbulence, damping constants, non-LTE, and 3D effects, which makes it difficult 
to calculate sufficiently accurate synthetic spectra.
Other problems are unidentified lines or lines with poorly known wavelengths
and $gf$-values.  Instead of a direct analysis,
low-resolution spectra may therefore be modelled based on a set of parameters and abundances (``labels")
available from a high-resolution study of a subset of stars.
A popular tool for this is {\em The Cannon}\footnote{The name refers to
Annie J. Cannon, who from 1911 to 1915 classified 225\,300 stars by 
visual inspection of Harvard objective prism photographic plates and later published the results as the 
Henry Draper (HD) catalogue} introduced by \citet{ness15}. It uses the labels of a learning
set of stars to find the parameters of a spectral model describing the flux 
in each spectral pixel as a function of the labels. A good example is provided by
\citet{ho17a, ho17b}, who apply {\em The Cannon} to transfer labels from APOGEE to 
$R = 1800$ LAMOST \citep{zhao12} optical spectra of K-giants using a learning set of $\sim 8000$ 
stars in common between the two surveys. This has resulted in metallicities \mh\ and
abundance ratios, \cm\ and \nm\ (derived from the CH and CN bands in the 4100 - 4400\,\AA\
region) for 450\,000 LAMOST giants with precisions $\simlt 0.1$\,dex relative
to the APOGEE values. Furthermore,
the ratio between $\alpha$-capture elements and iron, i.e.
\alphafe , is estimated to have a precision of $\sim 0.03$\,dex if the LAMOST spectra
have $S/N > 50$. A drawback of {\em The Cannon} is, however, that reliable
abundances cannot be determined for stars with atmospheric parameters and
compositions that are not well covered by the learning set of stars.

\subsection{Model atmosphere analysis, 3D and non-LTE effects}
\label{analysis}

In order to interpret observed stellar spectra in physical terms, models of the
light emitting regions, the stellar atmospheres, are needed. These are adequately
constructed by solving the conservation equations of mass, momentum and energy for
a compressible gas flow in the gravity field of the star, with a given energy 
input at the lower boundary, and radiative losses out to free space from the upper layers.
Particular care has then to be exercised in the treatment of
the radiative heating/cooling term in the energy equation. At every time step in 
the integration of the hydrodynamic equations, the equation of radiative transfer
has therefore also to be solved, together with the conservation equations,
for a considerable number of frequencies in order to represent the
total radiative flux properly.

Although a non-trivial task numerically, in particular due to the non-linear character of the
momentum (Navier-Stokes) equations, it is today quite possible to construct such models,
which may then for solar-type dwarfs be just a section of a stellar
atmosphere (a ``box in a star'') with adequate boundary conditions, 
where the time-development of the state of the gas in the box is simulated. When calculating the
integrated stellar spectrum, the spectra of a number of such states may be co-added to
represent the total flux from the star.
During the last 30 years, the methods to construct such 3D models have been successfully
advanced by Nordlund and collaborators \citep{Nordlund90a, Stein98, Asplund00a, Collet11, 
Trampedach13, magic13} and the resulting STAGGER code is now used for producing
extensive grids of models. Important and independent work along similar lines is also
made by a group behind the CO$^5$BOLD code \citep{Wedemeyer04, Ludwig09b, Freytag12}. 
In this latter group, also similar models of the full star (a ``star in a box'')
have been constructed for supergiants and AGB stars, where the photospheric granules are of sizes 
not very different from that of the full star so that ``box in a star'' models would be
inadequate.

At comparison with observations, these 3D hydrodynamic models have been quite successful.
The observed asymmetries of spectral lines, due to the convective motions of the gas,
 as a function of line depth and excitation were
reproduced \citep{Nordlund90b, Dravins90} and so was the variation of line strength with
transition probability (``the curve of growth'', \citet{Asplund00a}). 
The latter achievement was a major one from the perspective of abundance analysis, 
since it freed the analysis
from the suspect ``microturbulence velocity'', a parameter used to fit the strength of
partly saturated spectral lines in traditional studies (as mentioned already in Sect. \ref{param})
with no very clear physical definition. In fact, the
3D models showed that the observed line broadening ascribed to
``microturbulence'' was not mainly due to small-scale
turbulence but sooner the result of velocity gradients in large-scale granular
motions. One further impressive achievement of the 3D models was the demonstration that they
reproduced the various abundance measures for oxygen in the Sun, 
whether high-excitation or low-excitation atomic lines or molecular ones, with one 
consistent abundance, \citep[see][and references therein]{Pereira09}.
Other important achievements are the excellent agreement between predictions
from 3D models and observations of centre-to-limb variations of continuum intensity
\citep{pereira13} and iron lines \citep{lind17}.

Also, in another very important respect, the detailed 3D hydrodynamic models were able
to free the analysis from more ad-hoc parameters that are
as questionable as the microturbulence velocity. 
The traditional way of calculating the convective flux was namely
based on the ``Local Mixing Length Theory'', with roots in early work by 
Prandtl (1926, cited by \citet{Biermann32}), \citet{Biermann32}, and \citet{Vitense53}, 
which, when applied to stellar envelopes and atmospheres contains
several free parameters, notably the ``mixing length'' \citep{Henyey65, Gustafsson75} 
with definitions that are only partially possible to relate to the real physical situation. 
The hydrodynamic 3D
models are in practice not dependent on such parameters. One might ask whether the
maximum box size, or the minimum grid spacing, of the numerical representation of the
atmosphere would not after all be new significant free parameters.
Detailed comparisons with different spatial resolutions have shown, however, 
that this is not the case \citep{Asplund00b}.

When calculating the spectrum from the model atmospheres for comparison with observed
spectra, the interaction of the radiation with the gas
must be properly represented. In the general case this is a complex problem since
the number densities in the lower and upper states of a spectral line are affected by the
radiation, and moreover not only by the radiative field in the observed spectral line itself 
but by the populations in every state of that element (atom, ion, or molecule)
that directly or indirectly, through radiative or collisional transitions, 
significantly affect these states of the line. The fact that the real populations of the 
atomic states are not only the result of the statistical equilibrium of the
particular atom but also of the equilibrium of other atoms, ions and molecules,
for instance  of the relevant electron donors, means that in an abundance
analysis ``corrections for non-LTE'' must in principle be extended beyond the
study of individual atoms in a given model atmosphere. In order to
handle this problem, the full statistical-equilibrium equations 
have to be solved simultaneously for
every atomic state of significance and every spatial point in the model. In these equations,
the rate of transitions {\it to} any particular state is set equal to the rate of
 transitions {\it from} it. In this way, the number
densities of different atomic states are linked together in a large system of
linear equations, and moreover, also the radiative fields in the corresponding transitions
have to be considered when calculating the radiative transition rates.
Since the radiative fields are non-local, the radiative rates couple also
 the different spatial points in the model together.
The number of important states for an atom is excessive and the number of important transitions
(including both radiative and collisional ones) may be several thousands or more. Thus, this is clearly
a complex problem. It gets even more difficult since vast amounts of atomic data,
including radiative transition rates and collision transition rates due to
both atomic and electronic collisions, are needed and often missing.

For several decades, a major problem in statistical-equilibrium calculations for 
the modelling of stellar atmospheres and spectra has been the
the lack of realistic data for the cross sections of collision processes,
 both as regards excitation of atoms and molecules by collisions with
electrons and with hydrogen atoms. Until recently, only semi-empirical methods
 have been available which on good grounds were
expected to lead to collision cross sections that could be in error by even
 orders of magnitude. During the last decade, however,
this situation is greatly changed. Barklem, Belyaev and collaborators
have contributed realistic cross sections for hydrogen collisions with 
 atoms of Li, O, Na, Mg, Al, Si, Ca, K, Fe, and Rb, and cross sections for
electron collision excitation of Li, O and Mg, see the comprehensive
 review by \citet{Barklem16R}  and references therein. Most of these new data have now been
applied in spectrum analyses and greatly improved the validity of the results.
Of particular importance are the calculation of cross sections for 
hydrogen collisions with iron atoms by \citet{barklem18b}, already applied for a 
more accurate determination of the solar iron abundance by \citet{lind17}.

At present, relatively complete and accurate model atoms for calculations of
spectra from model atmospheres have been assembled for about 20 different atomic species.
However, for several of these model  atoms, important data like hydrogen collision 
cross sections and photoionization cross sections are only rough semi-empirical 
estimates that urgently need improvement.
Considerable effects on abundance determinations have been found
when applying statistical
equilibrium instead of older work in which Local Thermodynamic Equilibrium (LTE) was assumed
(that is that all energy and ionisation states were populated according to
Boltzmann-Saha statistics with one local depth-dependent temperature $T$).
Thus the ``non-LTE effects'' on abundances are often
significantly greater than the precision requirements discussed here of about 0.03\,dex
or better. 
In particular for metal-poor star where the over-ionization by the hot ultraviolet 
radiation may severely reduce the content of neutral atoms,
the estimates  by different groups of non-LTE effects may still depart significantly,
 e.g. reflecting differences in model atoms, atomic data, model atmospheres 
with different UV fluxes, etc. One example are the recent results by
\citet{shi18} and \citet{korotin18} for copper in metal-por stars
which differ considerably
probably as a result of the use of different cross sections. Similar uncertainties may plague
the analyses of several elements in the iron group such as Mn \citep[see e.g.][]{sneden16}. 
Critical comparisons of methods and results of different groups should routinely
be carried out.

Most of these statistical equilibrium calculations have, however,
 as yet been made only for plane-parallel models as discussed below.
 For 3D models, methods of different sophistication have been
applied in the radiative-transfer calculations. In the ``\lt 3D\gt approximation'', 
the physical properties at each horizontal layer in the 3D model is first averaged
before the spectrum is calculated. In the ``1.5\,D approximation'',
the radiative field is only integrated
in the vertical direction, while in ``full 3D'' departures from plane-parallel
stratification are considered in all directions. This step, to the ``full 3D'' (3D, below)
has been found to be significant in several cases \citep[see e.g.][]{amarsi16b,omalley17}, 
in  particular
for metal-poor models where the transparent gas may let radiation in
``side-ways'' from hotter regions, e.g. able to photo-excite/ionize the atoms of interest.
Obviously, the \lt 3D\gt models are not to be recommended for accurate analyses 
of very metal-poor stars, at least not for stars  close to the turn-off point 
where the non-local hot ultraviolet radiation may importantly affect the atoms in the cooler gas
sideways as well as vertically. Instead, in non-differential
analyses, full 3D models with spectrum calculations in statistical equilibrium 
should be used.

The model spectra described above have been calculated {\it a posteriori}, on the basis
of 3D models where the radiative field needed for the energy balance
equation of the model (as well as the thermodynamics of the gas) were
calculated assuming LTE. There is still some time before one
can expect fully consistent model atmospheres to be available where this LTE assumption
is lifted also at the calculation of model structure, and not only of the final spectrum.
 The most important steps towards consistency, seem, however, to have been taken:
 the modifications to expect from fully non-local radiation fields in 3D models are probably smaller.

Since 40 years, the model stellar
atmospheres used in most abundance analyses have, however, been far more primitive
and much less realistic than those described above: they have been based
on a number of simplifying and partly unrealistic assumptions:
plane parallel (or for larger stars spherically symmetric)
stratification of the gas in hydrostatic equilibrium,
LTE, and energy balance. Here, the radiative energy flux was calculated
 by the equation of radiative transfer with the Planck function $B_\nu(T)$ as
 the source function - if need be supplied with a scattering term. Added to that
 is the convective flux according to the
Local Mixing-Length ``Theory''. These 1D models have been calculated in extensive grids
extending over large ranges in fundamental parameters including chemical composition. The major
effort behind them has been directed towards the consideration of the
 collective effects of millions of spectral lines (``blanketing'')
 on the energy balance and the spectrum, and indeed, these
effects are very considerable on the temperature structures and the spectra,
 in particular for cool stars. The most
used grids of this type, that are still regularly applied in abundance analyses, are the ATLAS9 grid 
\citep{Kurucz79, Castelli03}, the MARCS grid \citep{Gustafsson75, gustafsson08} 
and the NextGen-grid \citep{Hauschildt99a, Hauschildt99b}. The latter models were calculated with the PHOENIX
programme which has also enabled the construction of models where the LTE assumption is not made.

The usefulness of the more primitive, and much more easily handled, 1D models in abundance analysis
has been explored in a number of
comparisons with results from 3D models. The early application to abundance analysis of 3D and non-LTE
modeling for solar-type stars was reviewed by \citet{Asplund05}. 
Until very recently, the application of 3D non-LTE models have mainly
been made for solar-type stars, including more metal poor ones with
temperatures and gravities not too different from the solar values. A general result is
that the differences from standard analyses are often greater than 0.04\,dex, and in particular for metal-poor
stars considerably greater than so. Thus, if the ambition is to determine abundances with
high accuracies, one should not use the 1D models without a very careful consideration of the
systematic errors involved.

\subsection{Differential analyses}
\label{differential}

One way of minimizing the effects of model errors is to perform a differential analysis relative
to the Sun, or some other standard star, hoping that the differences between the programme
stars and the standard star can be modelled accurately enough by 1D models. (In fact, a differential
analysis is anyhow recommendable in view of the fact, mentioned in the Introduction,
that in an absolute analysis, even for the Sun itself, it is as yet hard to achieve 
abundance errors smaller than 0.04\,dex.)

Schematically, we may then write any
estimate $ E$ of the logarithmic abundance ratio [A/B] for a programme star P,
determined differentially relative to a standard star S as
\begin{equation}
E[A/B]^P = [A/B]_a^P - [A/B]_a^S + [A/B]_r^S , 
\label{eqE}
\end{equation}
where an index $a$ denotes an approximate model (e.g., LTE and 1D) and $r$ a more realistic one, such
as a 3D model with statistical equilibrium calculations of the spectrum. A key question is then
how different the parameters of the programme star may be from that of the standard star if the
error in the estimated abundance ratio should be kept smaller than a certain value, say 0.04\,dex.
Another, related question is how the fundamental parameters of P and S are to be determined, in particular in
the fitting of the approximate models, for which the parameters in themselves as estimated
from the model spectra may not be correct {\it per se}.
Although these questions can certainly not be answered fully today for
stars located anywhere in the parameter space, there are enough of analyses for solar-like stars,
including metal-poor ones, to allow some conclusions.

In an important series of papers \citet{Bergemann12, lind12, amarsi16b, lind17},
developed model atoms for Fe I and Fe II 
with hundreds of energy level and thousands of radiatively-allowed transitions,
and applied these in 3D STAGGER models of six standard stars
 with different fundamental parameters, to study the effects on iron lines and resulting
iron abundances when realistic model atmospheres are used.
 We may consult these results to explore the errors of 1D LTE differential analyses
(note, here the word is not used  in a strict sense since different spectral lines
 will have to be used for stars with great abundance differences).
For the Sun and Procyon (the latter with estimated parameters
  $\teff = 6543 K$, log$g=3.98$ and [Fe/H]=-0.03) it is found that the
iron abundances are raised from standard 1D LTE values by typically 0.03\,dex for the Sun
 (dependent on which lines were used in the standard analysis) when  a \lt 3D\gt
 non-LTE analysis is carried out. For Procyon, the effect is an increase of about 0.06\,dex.
Hence, for the span of 700 K between Procyon and the Sun,
 a differential 1D LTE analysis  would still have worked
but marginally so, if the errors  required in  [Fe/H] are to be $< 0.04$\,dex.
A full 3D non-LTE analysis may in fact lead to smaller effects as 
suggested from the solar analysis by \citet{lind17}.
 It is worth noting that the differential effects for these stars, as long
as reliable \logg\ values are known,
are much smaller for Fe\, II lines than Fe\,I lines. 
Also, the errors in LTE analyses of \feh\ are partially compensated 
for if consistent LTE excitation equilibria are used for determining
the effective temperature.

\begin{table}
\centering
\caption{ The logarithmic iron abundances relative to the Sun, [Fe/H],
for four metal-poor stars derived in 1D, LTE analysis, and
obtained by \citet{amarsi16b} using STAGGER models and statistical-equilibrium
calculations of the spectra in 3D, respectively.
Given are also estimated statistical
and systematic errors in the latter [Fe/H] determinations.}
\label{tab:C}
\begin{tabular}{lrrrr}
\hline\noalign{\smallskip}
  & HD84937 & HD122563 & HD140283 & G64-12 \\
 \hline\noalign{\smallskip}
 $\teff $ & 6356 & 4587 & 5591 & 6435 \\
  \logg &  4.06 & 1.61 & 3.65 & 4.26 \\
 $ {\rm [Fe/H](1D, LTE)}$ & -2.10 & -2.60 & -2.42 & -3.24 \\
 ${\rm  [Fe/H](3D, non-LTE)} $& -1.96 & -2.49 & -2.31 & -2.94 \\
 $ \sigma({\rm stat})$ & 0.02 & 0.11 & 0.03 & 0.06 \\
 $\sigma({\rm syst})$ & 0.04 & 0.14 & 0.03 & 0.06 \\
  \noalign{\smallskip}\hline
  \end{tabular}
  \end{table}

In addition to the Sun and Procyon, four
metal-poor standard stars were discussed by the same authors.
 The model parameters of these stars are given in Table \ref{tab:C}.
As is seen, the errors in [Fe/H] in a differential analysis 
for the metal-poor stars will be considerably greater than those
discussed in this paper.
A closer look at the results of \citet{amarsi16b} demonstrates the intricacy
 of the interplay between the different effects. E.g., the iron abundance is raised
when non-LTE replaces LTE (mainly due to ``overionization'' of the neutral atom). If the 1D model is 
replaced by a \lt 3D\gt\ model, there is again a
small increase in the abundances derived, both if the spectrum is calculated in non-LTE 
 and if  LTE is assumed. However, if a full 3D model is used in the spectrum calculations,
the non-LTE abundances increase further while the abundances derived under
 the LTE assumption decrease significantly.

Iron is important, not only for its rich spectrum and relatively high abundance
 in solar-type stars. It is also one of the dominating electron donors, and
thus affects the continuous opacity through H$^{-}$.
 For the additional important electron donor in solar-type stars, Mg,
 \citet{bergemann17} found \lt 3D\gt non-LTE corrections of 1D LTE values of about 0.01-0.03 for
the Sun (depending on solar model and lines used) and corrections for Procyon of about 0.10.
Thus, errors on the order of 0.01 per 100 K
 will result in standard
analyses of solar-type stars relative to the Sun; however,
 just as for iron abundances it seems that most of those errors will be
 compensated for if iron-atom excitation equilibria from standard 1D LTE models
 are used for the temperature estimate. For the metal-poor stars,
 however, \citet{bergemann17} find corrections due to 3D that
contribute negatively to [Mg/H] while the non-LTE values increase the values
 -- altogether, these lead to positive corrections for stars like HD\,84937 and HD\,140283,
while the 3D effects dominate with negative corrections for HD\,122563.
 The errors may be up to 0.10\,dex in measures relative to the Sun.

In addition to these principal difficulties to bridge great abundance gaps
 in differential 1D LTE analyses, there is another more practical one: 
the spectral lines can be of such
different strengths for the standard stars, say the Sun, and the programme stars. Therefore,
 one generally has to take several steps via intermediate standards, using different sets of
spectral lines for determining the abundances of these relative to the primary standard
 and to the programme star. That introduces extra errors e.g. via the 
determination of the microturbulence parameter, which may require a more
detailed 3D non-LTE modelling also of the intermediate standard. In practice, it is therefore
 very difficult to reach accuracies better than 0.04\,dex relative to the Sun
for extreme metal-poor stars. However, it is quite feasible to determine
 precise abundances of these latter stars relative to another metal-poor star then used as a standard.

For silicon, also an important electron donor in particular for stars slightly hotter than the Sun,
statistical-equilibrium calculations by \citet{Bergemann13} and \citet{mashonkina16} suggest that
the differential non-LTE effects are very small for solar-type stars \citep[see also][]{nissen17}.
 We also note that \citet{amarsi17} find small 3D effects in the solar case, using
statistical-equilibrium calculations and 3D models.

Motivated by the significance of oxygen as the third most abundant chemical element
 in the universe, and its origin in massive core-collapse supernovae (SNe) with a short evolution
 time scale which makes it an important element in the study of galaxy chemical evolution,
considerable efforts have been made in exploring the errors made in oxygen abundance
 analyses with standard 1D models. From the comprehensive study with
a 22 level atom (43 radiative transitions), in 3D models, by \citet{amarsi16a},
 one finds that the relatively easily accessible 7773\,{\AA} triplet should not be used differentially in
1D LTE analyses across larger effective-temperature intervals than 50 K or 0.1\,dex in \logg ,
 if the systematic abundance errors are to be smaller than 0.04\,dex.
The alternative criteria, the high-excitation \OI  \,6158\,{\AA} line and 
the \oI \,6300\,{\AA }
 and  6363\,{\AA} forbidden lines are less problematic (although quite weak and therefore
difficult to measure at low and high
temperatures, respectively) but these lines are more sensitive to 3D effects
and difficulties in placing the continuum, and,
for the forbidden lines, plagued by blends, including the wide and shallow \CaI\
auto-ionization line across the 6363\,A line. If these blends can be controlled,
 the forbidden lines should lead to small errors, even across gaps as wide as 500 K
in \teff\ or 0.5\,dex in \logg\ for solar-type stars.

However, the situation may be more problematic for other chemical elements.
 One example is aluminium which has recently been studied with \lt 3D\gt\
models in non-LTE for a 42 level atom by \citet{Nordlander17}.
 While the differences between \lt 3D\gt\ non-LTE results and 1D LTE results are
small for the Sun as well as for the dwarf HD\,22879 with solar temperature but $\feh =-0.86$,
 and in fact also for the Pop I K0 giant Pollux, they increase to
0.4\,dex for the metal poor stars HD\,140283 and HD\,84937. Almost all this effect
 is due to overionization by the hot radiation released from deeper layers in
the transparent gas in the metal-poor stars.  This depletes the lower state of the
only useful line for metal-poor stars, the resonance line at 3961 {\AA}. 
For the more metal-rich stars several
 lines are possible to use, and this illustrates a problem which
certainly makes strictly differential analyses difficult when metal-poor stars are
 compared with much more metal-rich ones: the weights given to
different abundance criteria have in many cases to be shifted.
The 3D effects on the Al lines were considered after spatial and temporal averaging,
 except in the solar case where a more detailed treatment was carried out.
This, however, may not be very problematic since the 3D effects seem less important
 in this case than the departures from LTE.

 For sulphur, the errors in differential analyses should be small, according to the
 1D statistical-equilibrium calculations by \citet{takeda05b}. For calcium,  however, \citet{mashonkina17} has
 used 63 + 37 level model atoms for Ca I and Ca II respectively, allowing for
 charge-transfer and hydrogen-atom collisions rates calculated with cross sections from
 quantum-mechanical calculations by \citet{Barklem16}.  The error for Procyon
 when analysed in 1D LTE relative to the Sun is small for Ca I,
 while the abundance is found to be underestimated by 0.08\,dex when
 high-excitation Ca II lines are used. If metal-poor stars like HD\,84937
 or HD\,140283 are compared in a 1D LTE analysis
 with Procyon or the Sun, the systematic errors due to the LTE assumption
 will be typically $-0.2$\,dex.
 For titanium, the model atom used by \citet{bergemann11} has 216+77 levels for
 Ti\.I and Ti\,II, respectively, and more than 4000 transitions.
 The abundance effects in [Ti/H] for metal-poor stars like HD\,84937 and HD\,140283
 are typically underestimates by 0.15\,dex when LTE is assumed instead of non-LTE for Ti \,I,
 while they are close to zero for Ti\,II.  The corresponding absolute errors for the Sun
 are typically 1/3 to 2/3 as great, depending on the collision cross sections adopted.

In summary, we conclude that the systematic errors made in 1D LTE differential analysis admit
 spans between the programme stars and the standard star of
 about 500 K for some elements, provided that care is taken in selecting the abundance criteria.
 However, this is not possible for all elements. For Pop II stars the systematic errors in
 1D LTE analyses relative
to  Pop I stars with similar temperature and surface gravities are as a rule greater
 than 0.04\,dex, and the analyses must be corrected using 3D non-LTE models. Alternatively,
one may perform a standard analysis relative to another Population II star,
 and spend particular efforts in analysing that star with detailed models.
It should finally be noted that even a strictly differential LTE analysis between 
two stars with similar fundamental parameters may lead to significant systematic errors 
due to different non-LTE effects for elements with considerably different abundances 
in the two stars, especially if saturated lines are used.

\subsection{Effects of magnetic fields}
\label{magnetic}
In the discussion of 3D and non-LTE effects on abundance determinations
in the preceding sections we have neglected the possible effect of
magnetic fields. The quiet Sun is, however, known to be magnetic, and so are probably all stars.
The magnetic fields are expected to affect the stellar spectra, both through
the magnetic broadening of the spectral lines (``direct effects") and through forces
on the stellar plasma leading to different dynamic and
thermodynamic properties of the spectrum-forming gas (”indirect effects”).
\citet{fabbian12} explored these effects on the solar Fe spectrum, using
MHD simulations by the 3D Stagger code, starting from a uniform vertical magnetic field
configuration, and with LTE synthetic spectra with Zeeman splitting considered.
The spectra were compared  with results using 
non-magnetic models (i.e. pure HD simulations) with the same effective
temperatures and surface gravities,  and no Zeeman splitting. 
The authors found that the indirect effects are
generally the more important ones for neutral Fe lines, and may lead to an
underestimate of the solar Fe abundance by typically about 0.05-0.1\,dex.
This is due to the weakening of the lines in the line-forming regions caused by
the hotter gas in the magnetized atmosphere. The effect
is counteracted but not fully compensated for by the direct effects of
Zeeman broadening of the spectral lines.
In a subsequent paper \citet{fabbian15} studied the case of the \OI\ lines
and again found that an abundance determination using non-magnetic models would lead to underestimates
by a few to several centi-dex.

\citet{shchukina15} challenged the results of Fabbian et al. by a study of Fe spectra
from 3D models with and without magnetic fields assuming that the former are
dominated by small-scale dynamo action (with the net magnetic flux being zero).
Also for solar C, N, and O spectra the effects were found to be negligible \citep{shchukina16}.
This conflict was resolved by \citet{moore15}, who demonstrated that the assumed field
geometry is a key factor: a local-dynamo magnetic field gives one order of magnitude
smaller effects on abundances than a larger-scale vertical field.  Although the issue
concerning the true effects on photospheric spectra is not quite resolved,
the impressive success for 3D non-magnetic HD simulations in reproducing solar
observations has led \citet{pereira13} to suggest that the non-magnetic solar models
should be preferred.  It is not clear to which extent this conclusion is
valid for other types of stars.

We also note that the small corrections on derived solar
Fe abundances from including small scale magnetic fields with no net magnetic flux
are not totally negligible in discussions of very precise abundances of solar twins.
\citet{shchukina15} found an average correction of +0.016\,dex for \FeI\ lines,
but for low-excitation lines formed in the upper photosphere the correction reaches
$\sim +0.05$\,dex. Considering that the magnetic field configuration and strength may vary between
solar twins, there could be significant magnetic effects
on derived differential abundances.
Furthermore, the value of \teff\ derived from the excitation balance of \FeI\
lines may be affected, because the magnetic correction depends on excitation
potential. The correlation between Ba abundances derived from a 1D model
atmosphere analysis of the strong \BaII\
5853\,\AA\ line and the chromospheric activity for young solar twins
\citep{reddy17} may be connected to such effects, although the authors attribute
the correlation to the neglect of enhanced micoturbulence in the upper atmospheric
layers of active stars (see Sect. \ref{nucleo}).

Observational evidence for a connection between stellar activity and equivalent widths
of spectral lines have been known since long.  Recently, \citet{flores16} measured
the \CaII\ H\&K and H$\alpha$, H$\beta$ Balmer lines in about 300 high-S/N HARPS spectra
of the solar twin HD\,45184 observed from 2003 to 2014 and showed that this star with an 
age of $\simeq 2.7$\,Gyr has an activity cycle of 5.14\,yr.
Strong \FeII\ lines formed in the upper photosphere
were found to be modulated in their EWs over the activity cycle with an amplitude of $\sim 1.5$\,m\AA\
and there is also evidence for variations of \TiII\ and \BaII\ lines.
This suggests that at some level there are effects of chromospheric activity and
associated magnetic fields on derived abundances, but further studies are required
to see if this is a significant problem for obtaining abundances of solar twins
with  a precision of $\sim 0.01$\,dex.

\subsection{Cool stars}
\label{coolstars} 

The spectra of early and intermediate K stars are still dominated by atomic lines, 
and the continuum may be located in between the lines at least in the red and
near-infrared spectral region. Also, the modeling of the atmospheres of dwarfs, 
subgiants and giants of these types is nowadays fairly advanced 
(see in particular the 3D Stagger grid published by \citet{chiavassa18} and references therein). 
For these stars, with spectra of high quality, analyses of quite high precision 
relative to a proper standard star of similar spectral and luminosity type are possible 
and are encouraged.

The crowded visual …
spectra of late K, M and C stars, where the spectral 
continuum may be hard or impossible to locate, make however
high-precision abundance analyses more difficult.
 This problem can be handled for late K and early M stars by using red and infrared spectra
 where the continuum in places is still possible to trace. 
New detectors, spectrometers, models and not the least molecular and atomic data 
have led to great progress in this area \citep[for an early review, see][]{Gustafsson89} 
but it is still difficult to reach the high accuracy discussed in the present review.
 For still cooler stars, the millions of weak lines from polyatomic molecules,
 like H$_2$O for late M stars and HCN and C$_2$H$_2$ for C stars, change the continuum to an
 apparent quasi continuum. This requires analysis by synthetic spectra which is
 dependent on the availability of sufficiently accurate molecular data
(wavelengths, transition probabilities, dissociation energies and for 
statistical-equilibrium calculations, collision cross sections,
 photo-dissociation cross sections, etc.).
 Simultaneous determination of the abundances relevant for the molecular equilibrium
should be carried out. Another problem for M and C stars,
 which must be considered is the blanketing effects of the molecules
 on the atmospheres; thus, for these stars the abundance analysis and the model atmosphere
 construction should be done by iteration, or rely on interpolation in vast grids of
 models with different sets of abundance data.

Another matter of great significance in the analysis of red giant stars is the accurate 
determination of stellar fundamental parameters like effective temperature, and,
 in particular, surface gravity.
In a standard spectroscopic analysis of an early K giant, errors in abundances derived
 for neutral elements smaller than 0.04\,dex require errors in \logg\ smaller than
 0.08 to 0.20\,dex, depending on the element in question. This accuracy has been hard
 to achieve with pure spectroscopic methods. Now, however, with the possibility to use
 astero-seismological methods built on high-precision photometry
 \citep[e.g.][]{pinsonneault14, Takeda15}, including observations of convectively driven
 brightness variations which also scale with stellar surface gravity \citep{Kallinger16},
 surface gravities of much higher accuracy are possible to obtain for these stars.
 Given this possibility, one may now argue that at least for the numerous early-type K giants,
 many of which are He core-burning ``clump stars", it should be possible to reach a
 quite high abundance accuracy in differential analyses relative to a standard star
 of similar spectral type. For later K giants, and certainly for M giants,
 line-blending as well as continuum ``veiling" are aggravating the situation. 
Also, the model errors may be expected to be more severe, the later and thus
 higher up on the giant branch the stars are located,
where low surface gravities and low pressures enhance departures from LTE and 
spherical symmetry. These problems are also
illustrated by considerable and individual ”microturbulence parameters” which 
may lead to systematic errors for saturated spectral lines as long as realistic
 3D model atmospheres  are not used.

For cooler giants and AGB stars, the gravity estimates in the analyses have often 
been based on assumed absolute magnitudes and masses. In order to reduce errors
 in C or O abundances for carbon stars to less than about 0.04\,dex, one needs to
 know the parallaxes with an accuracy of about 10\%.
It remains to be seen how much {\em Gaia} parallaxes will improve this situation.
 Anyhow, uncertainties in model atmospheres is still a major worry for these stars,
 and hardly admit high accuracies, even if strictly differential approaches are attempted.
 Such analyses are problematic for these stars, in view of their pronounced individuality.

Pioneering studies of C, N, and O compositions of M stars were made by 
\citet{spinrad66}, \citet{lambert84}, and \citet{tsuji85}.
A first detailed study of CNO, Fe-peak and s-element abundances of M and MS stars was 
presented by \citet{Smith85}. It was based
on near-infrared and infrared high-resolution spectra, selected from wavelength regions
 only marginally affected by blending 
TiO bands. The signatures of dredge up of CN-processed material were shown,
 and enhancements of $s$-process elements shown for
the MS stars. The analysis was differential relative to the K giant $\alpha$\,Tau,
 and the errors in the abundances relative to
that star were typically estimated to 0.15\,dex.

\citet{Vieira86} pioneered the use the J band (1.1-1.4 $\mu$\,m) in analysis of M-star spectra 
and \cite{Onehag12} and \citet{Lindgren17}, see also \citet{Lindgren16}, performed
 high-resolution analyses of M dwarf spectra in this band, where the severe blending
 effects of TiO are limited. They have shown that a high internal accuracy can be achieved,
 and that the resulting metallicities agree relatively well for some M-type dwarfs
 in visual binaries with those of the hotter, and thus more safely analysed primaries.
 At least for earlier dwarfs than M4, errors in [Fe/H] smaller than 0.1\,dex are obviously
 possible. Similar work has been undertaken by \citet{veyette17}, who
estimated fundamental parameters, including \teff , \feh , and
\tife\ (\alphafe) for 29 M dwarfs from KeckII/NIRSPEC spectra at a resolution of 25\,000. 
These were calibrated semi-empirically using PHOENIX model atmospheres \citep{Allard12} 
and synthetic spectra, and, as regards the abundances, a set of wide 
binaries with M dwarfs and primaries of FGK-type. For the latter high-resolution
 spectra were analysed spectroscopically with standard methods. The basic assumption 
made was that the components in the binaries have identical chemical composition. 
The formal accuracies are estimated to be about 60\,K in \teff\ and 0.05\,dex 
in the abundance ratios, but more independent comparisons are needed to verify that 
the estimates are free from more systematic errors. 

Examples of such studies are
 those of \citet{ryde16} of 28 M-giants in the Galactic bulge, and \citet{Nandakumar18}
 of 72 Inner Galactic Bulge M giants where the high extinction also motivates studies
 in the K band at 2.1 $\mu$\,m. Here, Fe, Mg, and Si abundances were measured, although with
 typical errors estimated to be about 0.15\,dex. Yet another example is the study of
 abundances of M giants in the Galactic Center Nuclear Cluster by \citet{Do18} who found abnormal
 Sc, V, and Y abundances, again from K-band spectra. Their interpretation of the strong
 Sc I lines in terms of high abundances have, however, been questioned by \citet{thorsbro18}
 since even nearby M giants are found to show similarly enhanced K-band Sc I lines,
 suggesting that they are affected by departures from LTE.  Even for the warmer N-type carbon stars,
 detailed abundances are possible to derive from K and H band spectra \citep{Lambert86}.

\section{Stellar populations traced by chemical abundances}
\label{populations}
Like other spiral galaxies, the Milky Way consists of 
a bulge, a halo, and a disk. In order to understand the formation and evolution of
these components, it is important to know if each of them consists of stars with
a common origin (one population) or if they consist of several distinct populations. 
An important method to detect such sub-populations is based on high-precision
abundances, because it is likely that stars of a common origin have about the same
chemical composition. In principle, it should be possible to trace 
the origin of stars through ``chemical tagging" \citep{freeman02} or methods related
to phylogenetic studies \citep{jofre17a}.

Before 1950 it was generally believed that all stars had the same universal chemical composition,
but from analyses of  photographic high-resolution spectra in the 1950ties 
\citep{chamberlain51,baschek59,helfer59}, it was shown that
old globular cluster stars and high-velocity stars with halo kinematics 
have much lower iron-to-hydrogen ratios than the Sun and younger stars. 
This was interpreted as due to the synthesis of heavy elements by nuclear reactions
inside stars \citep{burbidge57} and dispersal of the products by mass loss
and supernovae explosions increasing the heavy element abundance 
of the star-forming interstellar gas with time, i.e. chemical evolution.
Furthermore, \citet{wallerstein62} showed that the abundances of Mg, Si, Ca , and Ti 
are enhanced relative to the abundance of Fe in metal-poor stars.
As these so-called $\alpha$-elements are promptly produced in massive stars and dispersed by
Type II SNe, whereas iron is also contributed by Type Ia SNe on a longer
timescale, \alphafe\ at a given [Fe/H] depends on the star formation rate 
in a stellar population \citep{tinsley79, matteucci86, gilmore98}. 
Therefore, \alphafe\ is an important label
of a stellar population, but other abundance ratios for which the production rates
depend on  stellar mass and types of supernovae or other sites for
nucleosynthesis, notably red giants with mass loss, are also important.

\subsection{Disk populations}
\label{disk}
The first major high-precision study of elemental abundances in disk stars was carried out by 
\citet{edvardsson93}, who selected F and G main-sequence stars in the solar neighbourhood
from the catalogue of $uvby$-$\beta$ photometry by \citet{olsen88} 
and divided them into nine metallicity
bins ranging from [Fe/H] $\sim -1.0$ to $\sim +0.3$ using the Str{\"o}mgren $m_1$ index. 
In each metallicity bin, high-resolution spectra with $S/N \simgt 200$ were obtained 
for the $\sim \! 20$ brightest stars and analysed with the EW method.
As a main result of the survey,
stars in the metallicity range $-0.8 <$ [Fe/H] $< -0.4$ were found to have 
differences in \alphafe\ correlated with their mean
orbital distance, $R_m$, from the Galactic center. 
For stars with $R_m < 7$\,kpc, [$\alpha$/Fe] was found to be on average about 0.15\,dex higher 
than \alphafe\ for stars with $R_m > 9$\,kpc. Assuming that $R_m$ is a measure of the
distance from the Galactic center of the stellar birthplace,
Edvardsson et al. explained the [$\alpha$/Fe] variations
as due to a star formation rate that declines with
Galactocentric distance; a higher [Fe/H] is reached in the inner parts of the 
Milky Way before Type Ia SNe start contributing iron.
This interpretation is, however, in disagreement with the vanishingly
small radial gradient in \alphafe\ found for open disk clusters and Cepheids
\citep[see][and references therein]{yong12, genovali15}.

\begin{figure}
\centering
\includegraphics[width=7cm]{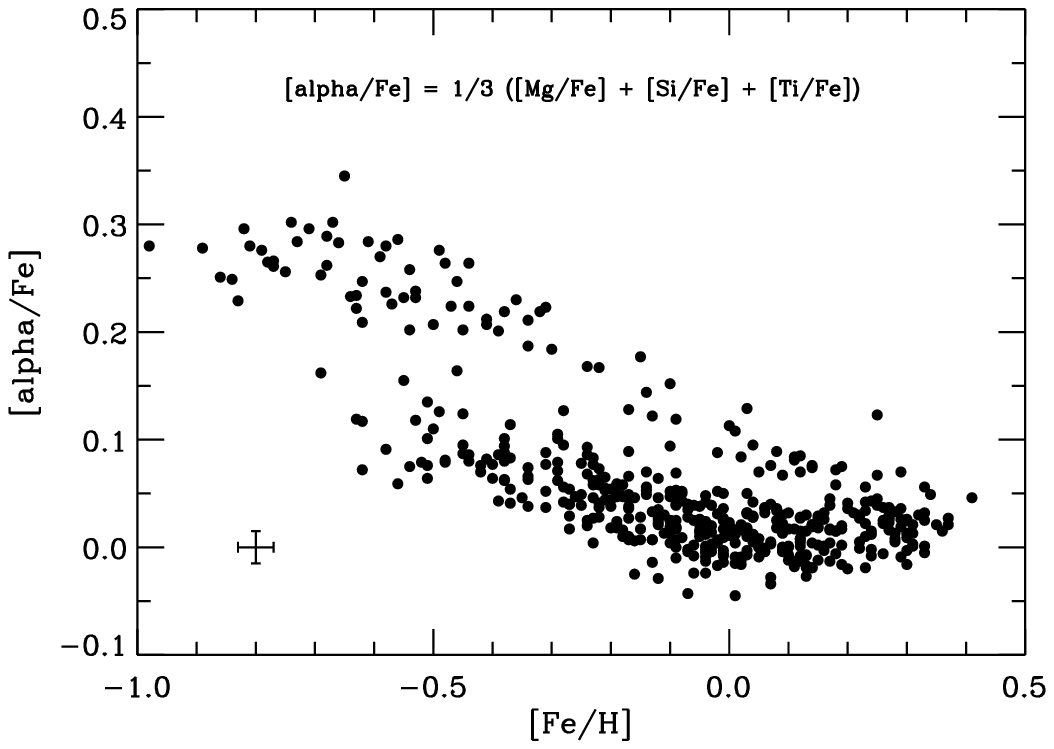}
\caption{\alphafe\ versus \feh\ for stars from \citet{adibekyan12} selected to have
\teff\ values  within a range of  $\pm 300$\,K from \teff\ of the Sun.
\cafe\ is not included in the definition of \alphafe , because the splitting
between the high- and low-alpha sequences is about a factor of two smaller in \cafe\
than in \mgfe , \sife , and \tife . Based on data from \citet{adibekyan12}.}
\label{fig:adibekyan}
\end{figure}

\begin{figure}
\centering
\includegraphics[width=12cm]{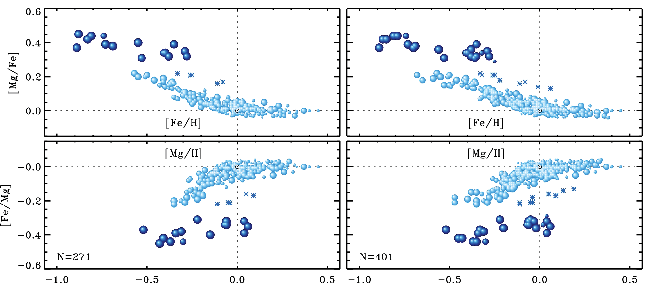}
\caption{\mgfe\ versus \feh\ (upper panels) and \femg\ versus \mgh\ (lower panels)
for a volume complete sample of F and G stars within a distance of 25\,pc.
The left panels refer to a northern sample
($\delta > -15^{\circ}$) discussed by \citet{fuhrmann11}, whereas the right panels
show an all-sky survey, where the $d < 25$\,pc selection is based on the revised
Hipparcos distances of \citet{vanleeuwen07}.
Stars on the high-Mg sequence are shown with dark blue circles
and stars on the low-Mg sequence with light blue circles. The size of the circles is
proportional to the stellar age. High-Mg stars have ages around  12\,Gyr,
whereas the ages of low-Mg stars range from 1 to 8\,Gyr. Stars lying
between the high- and low-Mg sequences shown with asterisks have ages around 10\,Gyr.
Reproduced from \citet{fuhrmann17} with permission from Klaus Fuhrmann.}
\label{fig:fuhrmann}
\end{figure}

The variations in \alphafe\ for metal-poor disk stars were confirmed by independent
high-resolution studies of \citet{fuhrmann98} and \citet{gratton00}. They interpreted, however,
the results as a dichotomy in \alphafe\ between stars having thin- and thick-disk 
kinematics\,\footnote{As shown by \citet{gilmore83}, the 
distribution of stars as a function of distance from the Galactic plane 
suggests the existence of two
distinct components: the thin disk with a scale height of 300\,pc corresponding to
a velocity dispersion $\sigma (W) \simeq 20$\,\kmprs\ and the thick disk having a
scale height of 1300\,pc and $\sigma (W) \simeq 40$\,\kmprs .}. This interpretation was supported by
\alphafe\ values for kinematically selected samples of thin- and thick-disk stars in
the solar neighbourhood by \citet{bensby05} and \citet{reddy06}, but it was unclear from these studies,
if stars with \alphafe -values between those of the thin and thick disk exist. 
Clarification has come from two important high-precision studies 
of samples of nearby stars selected without any kinematical bias. \citet{adibekyan12}   
derived abundances of 13 elements from HARPS $R = 115\,000$ spectra of a magnitude limited
sample of 1111 FGK main-sequence stars and \citet{fuhrmann17} used $R = 50\,000$ spectra to
obtain Mg and Fe abundances for  
$\sim 500$ dwarf and subgiant stars with $\teff > 5300$\,K within a distance of 25\,pc.  
While Adibekyan et al. derived
abundances with the EW method, Fuhrmann et al. made spectrum synthesis fits to
selected lines including Balmer lines for \teff\ determination and
the Mgb triplet as a constraint on gravity.
For both samples, the stars
split up into two sequences, high- and low-alpha stars, respectively, 
clearly separated in \alphafe\ for the metallicity range $-0.7 < \feh < -0.3$
as seen from the diagrams shown in Figs.\,\ref{fig:adibekyan} and \ref{fig:fuhrmann}.
A dichotomy is also seen in the $\ofe - \feh$ diagram
of \citet[][Fig.\,1]{bertrandelis16} obtained from APOGEE abundances of K giants in the 
temperature range $4200 < \teff < 4600$.

As suggested by \citet{fuhrmann17}, there may be a distinct population of stars
with intermediate values of \mgfe\ shown by asterisks in  Fig.\,\ref{fig:fuhrmann}.
Similar stars lying above the 
low-alpha sequence in the metallicity range $-0.3 < \feh < +0.2$ were found
by \citet{adibekyan12} (see  Fig.\,\ref{fig:adibekyan}) and interpreted by them as a special
high-alpha metal-rich (``h$\alpha$mr'') population in the solar neighbourhood,
perhaps coming from the inner regions of the Milky Way. However, they may just be an extension of 
the high-alpha sequence as suggested by  \citet{bensby14} based on their   
\tife -\feh\ diagram of kinematically selected stars in the solar neighbourhood. 
Such stars are also seen in the Galactic bulge \citep[e.g.][]{bensby17}.
Further studies are needed to clarify the origin of these alpha-enhanced stars
in the solar neighbourhood with metallicities around or above solar metallicity.

For the stars in Fig.\,\ref{fig:fuhrmann}, \citet{fuhrmann17} have derived
ages from evolutionary tracks in the $M_{\rm bol}$-\teff\ diagram.
The high-Mg stars are all very old ($\sim 12$\,Gyr), whereas the ages of 
low-Mg stars range from 1 to 8\,Gyr
and the intermediate stars have ages around 10\,Gyr. These results agree with ages derived
for the Adibekyan et al. sample by \citet{haywood13}, who  consider the
intermediate stars as an extension of the high-alpha sequence formed 
in a ``thick-disk"\,\footnote{This designation
is somewhat misleading, because the majority of the metal-rich stars on the high-alpha sequence
do not have thick-disk kinematics} phase of the Milky Way with intense star formation
lasting about $\sim 4$\,Gyr and with a downward trend of \alphafe\ as a function of time
due to an increasing contribution of iron from Type Ia SNe. After a quenching of the star 
formation for a period of $\sim 1$\,Gyr, 
metal-rich thin-disk stars began to form $\sim 8$\,Gyr ago \citep{snaith14} 
leading to a slight decrease of \alphafe\ with time as also found for solar twins
\citep{nissen15, spina16a}.
The metal-poor low-alpha stars are, on the other hand, interpreted 
as a distinct population born in the outer disk
and mixed to the solar neighbourhood by orbital diffusion, whereas the majority of
high-alpha stars have been born in the inner disk as suggested from
their rotational lag relative to the Local Standard of Rest (LSR). This scenario 
is supported by in-situ abundances from APOGEE \citep[][Fig.\,4]{hayden15};
the high-alpha sequence is very pronounced in the inner Galaxy but almost disappears
in the outer disk, where the low-alpha sequence dominates.

\subsection{Halo populations}
\label{halo}
For a long time it has been discussed if the Galactic halo consists of more than 
one population of stars. The answer to this question is important for understanding
how the Milky Way formed. The monolithic collapse model of 
\citet{eggen62} corresponds to a single old population, but from a study of globular clusters
\citet{searle78} suggested that the halo
consists of two populations, i.e. an inner population formed in-situ during a dissipative
collapse and an outer spherical population accreted from dwarf galaxies.

Correlations of \alphafe\ with the kinematics of halo stars
have given support to the Searle-Zinn scenario 
\citep{nissen97, fulbright02, stephens02, ishigaki10}, but the first
clear evidence for the existence of two discrete halo populations 
was found by \citet{nissen10, nissen11}.
Based on the $uvby$-$\beta$ catalogue of \citet{schuster06},
they selected a nearby ($d < 340$\,pc) sample of 94 F and G main-sequence stars 
with $-1.6 < \feh < -0.4$ of which the majority have halo kinematics, 
i.e. a total space velocity with respect to the LSR, 
$\Vtotal > 180$\,\kmprs, while 16 stars have thick-disk kinematics.
High-resolution, $S/N \sim 200$ spectra were obtained with the ESO VLT/UVES
and NOT/FIES spectrographs and used to derive atmospheric parameters
and differential abundances of
Na, Mg, Si, Ca, Ti, Cr, Mn, Fe, Ni, Cu, Zn, Y, and Ba using the EW
method in the LTE approximation. The differential errors of \xfe\ range from 0.01 to 0.04\,dex depending 
on the number and strength of lines available for a given element. 
In addition, \citet{ramirez12} and \citet{nissen14} have derived C and O abundances for the 
same sample of stars taking into account non-LTE effects.

\begin{figure}
\centering
\includegraphics[width=11cm]{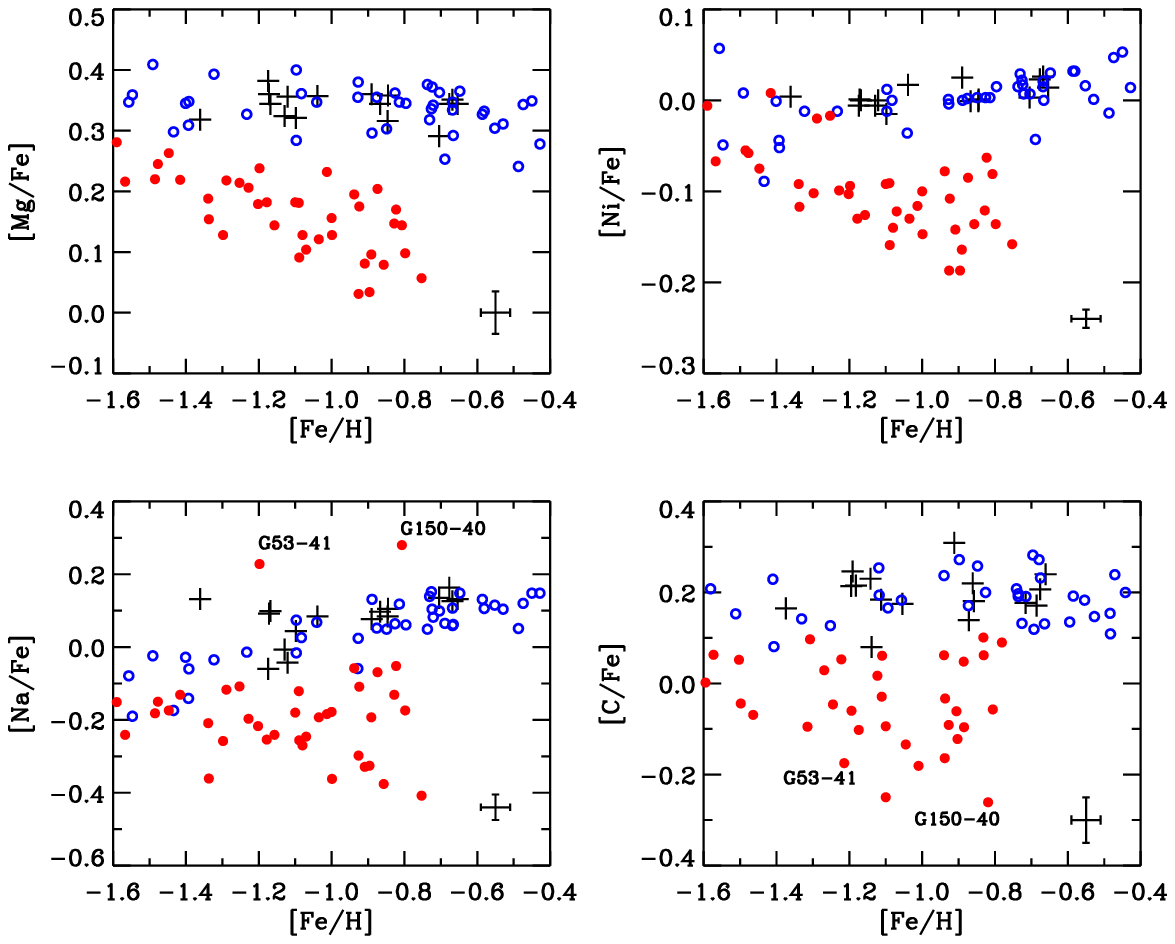}
\caption{Abundance ratios versus \feh\ for the halo and thick-disk stars in 
\citet{nissen10}. Stars with halo kinematics
have been divided into high-alpha stars shown with open blue circles and low-alpha stars
shown with filled red circles. Stars with thick-disk kinematics are shown with crosses.
Two stars with peculiar high \nafe\ values are marked. They have low \cfe\ and \ofe\ 
values and may be born as second generation stars in globular clusters.}
\label{fig:NS10}
\end{figure}

As seen from Fig.\,\ref{fig:NS10}, the \mgfe -\feh\ diagram of \citet{nissen10}
suggests that stars with halo kinematics can be divided into two separate populations, 
i.e. a high-alpha sequence for which \mgfe\ lies around a plateau of
$\sim 0.35$\,dex like the  thick-disk stars and a low-alpha sequence for which \mgfe\
declines with increasing metallicity. The two populations merge at $\feh \simeq -1.6$
but have a separation in \mgfe\ of about 0.2\,dex at $\feh = -0.8$; 
interestingly, there are no low-alpha stars with 
$\feh > -0.7$. A separation of the two halo populations is also seen for other abundance ratios
such as \cfe , \ofe , \nafe , \sife , \tife , \nife , \cufe , \znfe , and \yba , whereas there is no
or very little separation in \cafe , \mnfe , and \crfe . 

APOGEE abundances of K-giants with $-1.4 < \feh \ -0.7$  and
distances reaching beyond the solar neighbourhood
confirm the existence of two distinct halo populations
\citep{hawkins15, hayes18}. In addition to differences in \cfe , \ofe , \mgfe , and \nife ,
the APOGEE data also show a very clear difference in \alfe , a ratio which was not explored
by \citet{nissen10}. 

The trends of \alphafe , \nafe , \nife , \znfe , and \yba\ as a function of \feh\ for
the low-alpha population resemble the corresponding
trends for giant stars in dwarf spheroidal (dSph) galaxies such as Fornax and Sculptor 
\citep[e.g.][]{shetrone01, shetrone03, tolstoy09, kirby09, letarte10, skuladottir17}. 
This has led to the suggestion that
the low-alpha stars originate from dwarf galaxies
with relatively low star formation rates, \citep[see the
pioneering paper by][]{matteucci90},
allowing Type Ia SNe to start contributing iron at metallicities already around
$\feh \sim -1.6$, whereas high-alpha stars have formed 
in halo regions with such a high star formation rate that Type Ia SNe
did not contribute iron until $\feh \sim -0.4$.  For a detailed discussion of this
scenario and effects of variations in the initial mass function,
we refer to \citet{mcwilliam13} and \citet{fernandez-alvar18}. 

\begin{figure}
\centering
\includegraphics[width=8cm]{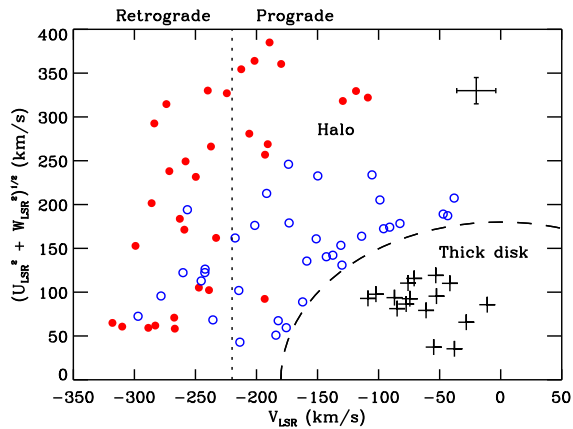}
\caption{Toomre diagram for halo and thick-disk stars  
in \citet{nissen10} shown with  the same symbols as in Fig.\,\ref{fig:NS10}.
Only stars with $\feh > -1.4$ are included.
The long-dashed line corresponds to $V_{\rm total} = 180$\,\kmprs\
and the dotted line separates stars moving in retrograde orbits from stars
with prograde orbits.}
\label{fig:toomre}
\end{figure}

Further evidence that the low-alpha stars  belong to an accreted population
is obtained from their kinematics. As seen from the Toomre diagram in
Fig.\,\ref{fig:toomre}, they tend to move on retrograde orbits and
have on average larger $U$ and $W$ velocities than the high-alpha stars.
This is confirmed by recent studies of APOGEE abundances in combination
with kinematical data from the Gaia DR2 release \citep{gaia.collaboration18}.
As suggested by \citet{haywood18} and \citet{helmi18} most of the
low-alpha stars may be due to accretion of a single massive satellite galaxy.
Evidence of this event is seen as a prominent slightly retrograde
structure in the Toomre diagram
of $\sim 6000$ high-velocity stars ($\Vtotal > 210$\,\kmprs ) with distances less than 1\,kpc
\citep{koppelman18}.

\begin{figure}
\centering
\includegraphics[width=12cm]{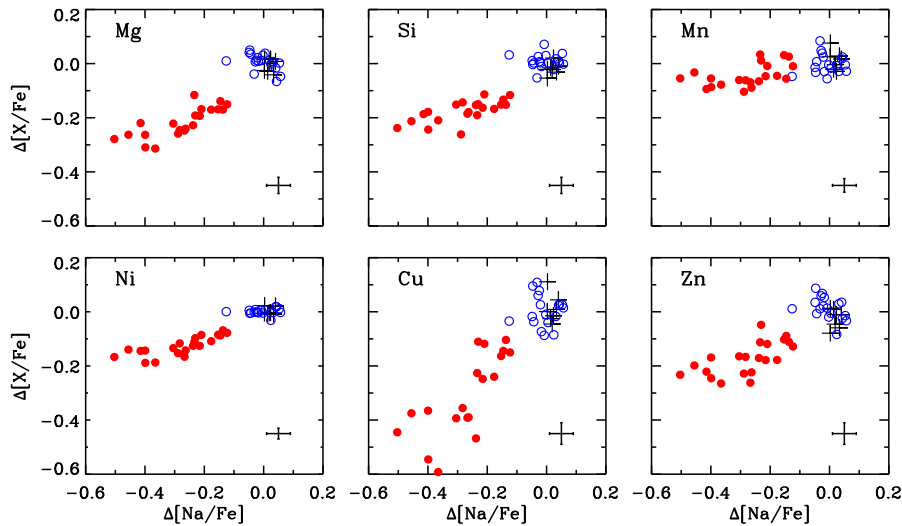}
\caption{Correlation between differential abundance ratios for halo and thick-disk stars
in \citet{nissen10, nissen11} shown with the same symbols as in Fig.\,\ref{fig:NS10}.
Only stars in the metallicity range $-1.1 < \feh < -0.7$ are included.}
\label{fig:NS11}
\end{figure}

As seen from Fig.\,\ref{fig:NS10}, the dispersion of abundance ratios \xfe\
for stars on the low-alpha sequence 
in the metallicity bin $-1.1 < \feh < -0.7$ is a factor 2\,-\,3 higher than the dispersion for 
high-alpha and thick-disk stars. This additional dispersion among the low-alpha stars cannot
be explained by errors in the abundance ratios;
the high- and low-alpha stars have about the same distribution  in \teff , \logg ,
and \feh , so non-LTE corrections are expected to change the mean abundances and 
dispersions for the two populations in a similar way. This has been confirmed in the case
of Cu from a non-LTE analysis of the \citet{nissen10} spectra by \citet{yan16}. 

Figure \ref{fig:NS11} shows that there is as strong correlation between 
$\Delta \xfe$ (measured relative to the average \xfe\ for the high-alpha population in
the $-1.1 < \feh < -0.7$ bin) and $\Delta \nafe$ , which is the abundance ratio
showing the largest variation. Stars in dSph galaxies have
more extreme values of $\Delta \xfe$,
see e.g. the \nife \,-\,\nafe\ relation for Fornax by \citet[][Fig.\,13]{letarte10}.
This suggests that the low-alpha stars did not form
in systems like present-days dSph galaxies, but they may originate from more
massive satellite galaxies as proposed by \citet{zolotov09, zolotov10}. Based on   
$\Lambda$CDM modelling of galaxy formation, they  predict a dual
origin of stellar halos. High-alpha stars were formed in the high-density inner $\sim 1$\,kpc
region but were displaced to the halo by mergers of a few massive 
satellite galaxies in which the low-alpha stars were born and later accreted to the halo.
According to this, the dispersion in \xfe\ for the low-alpha population is 
due to differences in the mass of the satellite galaxies, i.e. a lower mass 
corresponds to a lower star formation rate or, alternatively, a higher mass-loss rate
from the system, and hence a larger deviation of \xfe\
from the value on the high-alpha sequence. If only a few dwarf galaxies 
contributed to the low-alpha population, there could be sub-structure in the  
abundance ratios. A hint of this may possible be seen in the $\Delta \cufe$\,-\,$\Delta \nafe$
diagram, but a much larger sample of stars with very precise abundances is needed
to verify the presence of sub-populations.

\subsection{Extremely metal-poor stars}
\label{extreme}
The two distinct populations discussed in the previous section have been
detected among the more metal-rich halo stars, i.e. in the metallicity range
$-1.5 < \feh < -0.7$. Obviously, it would be interesting if a similar dichotomy 
in the distribution of abundance ratios is present for more metal-poor halo stars
or if there is a cosmic scatter in abundance ratios at a given metallicity
as predicted by some stochastic models 
\citep[e.g.][]{argast00, karlsson05, bland-hawthorn10} dealing with the 
chemical enrichment at extremely low metal abundances ($\feh < -3.0$),
where only a few supernovae have contributed to the composition of 
the low-mass stars that can be observed today. 

\citet{frebel15} have given a comprehensive review of chemical abundances 
in extremely metal-poor stars with emphasis on the existence of carbon-normal
and carbon-enhanced stars as evidence of two distinct channels of star
formation at the earliest times. Here we limit our review to a few works,
where the precision of the derived abundances is approaching 0.03\,dex.

In the metallicity range $-2.8 < \feh < -1.5$, abundances
of 18 elements, from Li to Ba, have been obtained by \citet{reggiani17}
for a sample of 23 main-sequence stars selected to have effective temperatures
in a range from 6000\,K to 6500\,K.
VLT/UVES spectra with a resolution of $R \sim 50\,000$ and $S/N \sim 200$
were analysed with the EW method assuming LTE but adopting non-LTE corrections for
\NaI , \AlI , and \CrI . There is no indication of a dichotomy in the 
distribution of abundance ratios like in the case of more metal-rich halo stars
and for the $\alpha$-capture and iron-peak elements the standard deviations of 
\xfe\ relative to straight line fits as a
function of \feh\ are not significantly larger than the estimated abundance errors, i.e.
0.04 to 0.08\,dex. This is in agreement with stochastic models
predicting negligible cosmic scatter among halo stars with $\feh > -2.5$
due to mixing of products from hundreds of SNe. However, for the neutron-capture elements, 
Sr, Y, and Ba, there appears to be a significant cosmic scatter on the order
of 0.20\,dex.

At still lower metallicities, $-4.0  < \feh < -2.7$, \citet{cayrel04} have derived
abundances of 17 elements, from C to Zn, for 35 giant stars of which 22 have
$\feh < -3.0$. UVES spectra with a resolution of $R \sim 45\,000$ and 
$S/N \sim 100$ to 200 were obtained and analysed with the EW method
in LTE but applying 3D corrections for the \oI\ 6300\,A line and non-LTE corrections
for the resonance lines of Na, Al, and K. 
The scatter for most elements around fitted straight
lines, $\xfe = a \feh + b$, is of the same order as the measurement 
errors, i.e. 0.05 to 0.10\,dex, except for C and Na. In the case of \mgfe , a
small scatter, i.e. 0.06\,dex, is confirmed in a high-resolution study of 
23 main-sequence turnoff stars having $-3.4 < \feh < -2.2$ by \citet{arnone05}.
These results do not agree with the stochastic chemical enrichment models
of \citet{argast00} and \citet{karlsson05}, who predict $\sigma \mgfe \sim 0.2$ to 0.4\,dex 
(depending on the adopted yield of core-collaps SNe) for the metallicity range   
$-4.0 < \feh < -3.0$.

In contrast to the small scatter of \xfe\ for the $\alpha$-capture and
iron-peak elements, there is a large scatter for the neutron-capture elements 
among stars with $\feh \simlt -3.0$. For the \citet{cayrel04} 
sample of K giants, \citet{francois07} find a scatter of 0.6\,dex for \srfe\ 
and for Y, Zr, and Ba the scatter of \xfe\ is $\sim 0.4$\,dex, i.e. significantly 
greater than the observational errors of $\sigma \xfe \sim 0.1$\,dex. Also for
heavier elements like Eu, Gd, Dy, and Er there is a cosmic scatter of \xfe .
As discussed by e.g. \citet{cescutti15}, this may be explained by stochastic
chemical evolution models with contributions to the $r$-process from both
core-collapse SNe and merging neutron stars. In view of the gravitational-wave 
detection of two merging neutron stars by LIGO/Virgo \citep{abbott17} and the likely
production of $r$-process elements in connection with this event
\citep[e.g.][]{cote18}, it would be interesting to obtain
high-precision  abundances of neutron-capture elements in a large sample of extremely
metal-poor stars to test the relative importance of neutron-star mergers and
core-collapse SNe as sites for the $r$-process.   

\begin{figure}
\centering
  \includegraphics[width=8.0cm]{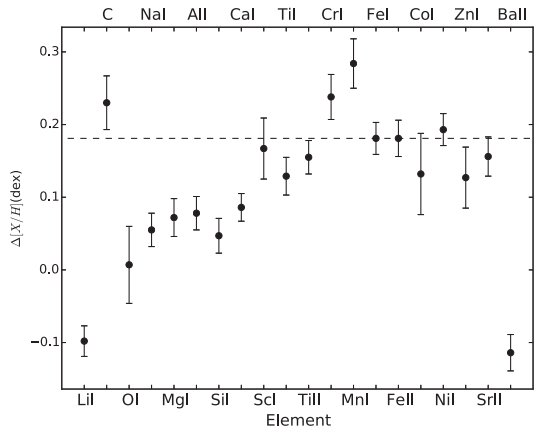}
\caption{Differences in abundances between the two extremely metal-poor stars,
G\,64-37 and G\,64-12.
Reproduced from \citet{reggiani16} with permission from Henrique Reggiani.}
\label{Fig4.Reggiani}
\end{figure}

As discussed by \citet{karlsson05} there is special reason to obtain high-precision 
abundances for very metal-poor halo stars, namely to look for narrow Single Supernovae Sequences 
(SSSs) in various 
abundance diagrams, e.g. \timg\ versus \camg , representing the location of stars
with major contributions from single SNe of different masses.
If such SSSs were detected one would get direct information about first generation supernovae yields.
The prediction of these sequences is based on the assumption that star formation
is unclustered and that stars are randomly distributed in space. If instead long-lived
stars in the halo are primarily formed in massive clusters with a homogeneous
composition, one should see tight groups of stars in abundance space at $\feh < -3.0$
\citep{bland-hawthorn10}. In both cases the challenge will be to obtain abundances 
with precisions better than 0.05\,dex for hundreds of very metal-poor stars.

\citet{reggiani16} have shown that it is indeed possible to reach precisions of
$\sigma \xfe < 0.05$\,dex for the relative abundances of very metal-poor stars.
Spectra with $R \sim 95\,000$ and $S/N \sim 800$ for two main-sequence turnoff 
stars (G\,64-12 and G\,64-37 with respectively $\feh = -3.2$ and $-3.0$)
were obtained with HIRES at the Keck 10\,m telescope
and analysed in a strictly (line-by-line) differential way assuming LTE. 
Interesting differences of \xfe\ between the two stars have been found
(see Fig. \ref{Fig4.Reggiani}) suggesting that SNe of different masses have enriched
the gas clouds out of which the stars were formed. Given that G\,64-12 and G64-37
($V = 11.45$ and $V = 11.15$) are among the brightest stars having $\feh < -3.0$,
one probably needs observing time at extremely large telescopes
to obtain spectra with similar high quality for hundreds of extremely metal-poor stars,

\subsection{Bulge populations}
\label{bulge}
It has been much discussed whether the Milky Way has a ``classical" bulge consisting
of old stars formed rapidly during an initial dissipative collapse phase or if it 
is a ``pseudo-bulge" formed over a longer period due to dynamical instabilities
in the Galactic disk. Stellar abundances may help to answer
this question by providing information on the timescale for the formation of
the bulge. 

Beginning with the pioneering model-atmosphere analysis of high resolution
spectra of K giants in the Galactic bulge by \citet{mcwilliam94}, 
there have been many studies of 
the \alphafe\,-\,\feh\ trends with conflicting results \citep[see review by][]{mcwilliam16}.
In some works a high plateau of $\alphafe \sim 0.3$ extends to solar metallicities
corresponding to a very high star formation rate as predicted for models of a classical bulge.
In other works \alphafe\ starts to decline at $\feh \sim -0.5$ like
in the thick disk supporting the pseudo-bulge scenario. These differences are probably
due to statistical and systematical errors of \feh\ and 
\alphafe ; as shown by \citet{schultheis17}
there are differences on the order of 0.10 to 0.15\,dex between previously
published abundances and new abundances derived from APOGEE spectra. 

The median \mgfe\,-\,\feh\ relation based on 
APOGEE abundances for $\sim 300$ bulge giants in Baade's window
\citep[][Fig.\,11]{schultheis17} 
is nearly the same as the relation for local thick-disk stars, 
but there is a large scatter in \mgfe\ at a given \feh ; for stars with
$\feh > 0.0$, \mgfe\ ranges from $-0.05$ to $+0.25$\,dex.
It is unclear if this scatter is cosmic or due to errors in \mgfe . A more tight
\mgfe\,-\,\feh\ relation was found for $\sim 2000$ red clump giants in 11 bulge
fields by \citet{rojas-arriagada17}, who used $R \simeq 16\,000$, $S/N > 80$ 
spectra of the  8480\,-\,9000\,\AA\ region  from the {\em Gaia}-ESO survey
to derive atmospheric parameters and abundances. Except for a few outliers
with low \alphafe\ values \citep{recio-blanco17}, the \mgfe\,-\,\feh\ relation
agrees well with that of local thick-disk stars. Interestingly, the metallicity
distribution of the bulge stars is bimodal with peaks at $\feh \sim -0.4$ and
$\feh \sim +0.3$ and with the metal-rich population claimed to reach \feh\ values as high as
+0.7\,dex. Based on the kinematics as determined from radial velocities,
\citet{rojas-arriagada17} associate the metal-rich population with a boxy/peanut bulge 
formed in the early thin disk and the metal-poor
population with a classical bulge, although it
could not be ruled out that it is due to a secular evolution of the early thick disk.

During the last decade, \citet{bensby17} have carried out a unique study of the 
Galactic bulge by determining abundances and ages of microlensed
dwarf and subgiant stars. Normally such stars in the bulge
have apparent $V$ magnitudes in the range 18\,-\,20, but during a microlensing event
they brighten to $V = 13 - 16$ for a few hours. This made it possible to
obtain high-resolution spectra  ($R \sim 40\,000$) with S/N up to 200
for 90 stars primarily with the UVES spectrograph at the ESO VLT. 
The spectra were analysed with the EW method in the same way as spectra of 
thin- and thick-disk stars in the solar neigbourhood \citep{bensby14}. 
Abundances of 11 elements (Na, Mg, Al, Si, Ca, Ti, Cr, Fe, Zn, Y, and Ba)
were determined with errors of \xfe\ typically in the range 0.05\,-\,0.10\,dex.
Furthermore, ages could be estimated from isochrones in the \logg\,-\,\teff\ diagram.

\begin{figure}
\centering
  \includegraphics[width=12.0cm]{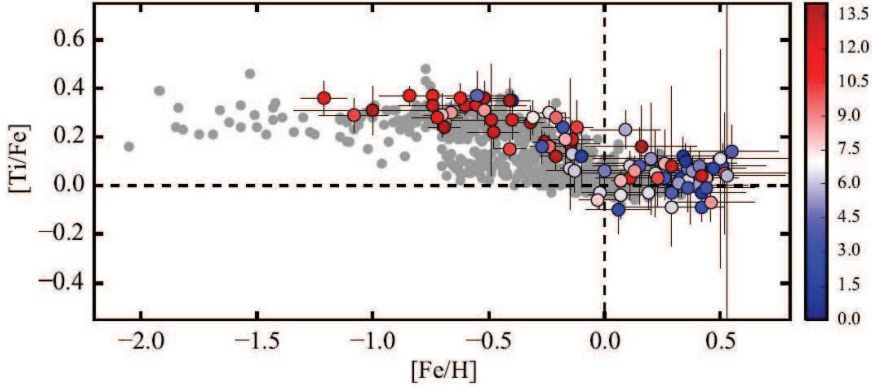}
\caption{\tife\ versus \feh\ for microlensed dwarf stars in the Galactic bulge.
The colours show isochronic ages in Gyr as given in the bar on the
right-hand side. Grey circles are the solar neighbourhood dwarf stars from
\citet{bensby14}. Reproduced from \citet{bensby17} with permission from Thomas Bensby.}
\label{Fig20.Bensby}
\end{figure}

The \xfe\,-\,\feh\ relations of the microlensed stars
follow the trends of local thick-disk stars,
but with a $\sim 0.10$\,dex higher \feh\ of the ''knee", where \alphafe\ starts to
decline from the plateau value, see Fig. \ref{Fig20.Bensby}.
There are indications of several peaks in the 
\feh\ and age distributions and about 35\,\% of the bulge stars
with $\feh > 0$ have ages younger than 8\,Gyr. This seems at odds with
colour-magnitude diagrams suggesting that all bulge stars are very old. However,
as pointed out by Bensby et al., old metal-poor and young metal-rich
isochrones are nearly indistinguishable
from each other. Altogether, the results of Bensby et al. favor a secular origin
of the Galactic bulge, but a minor classical component cannot be excluded.

Further improvements of the precision of abundance determinations for bulge stars
to a level of $\sim 0.03$\,dex would be important to obtain more information about
multiple populations in the Galactic bulge. Possibilities of using infrared spectra
of late K and M giants for this purpose were discussed in Sect. \ref{coolstars}.

\subsection{Chemical tagging and phylogenetic studies}
\label{tagging}
The idea of ``chemical tagging" is to use stellar abundances
to identify stars now widely dispersed in space as born in the same molecular cloud. It is based
on the assumption that stars in a given cluster
have the same chemical composition at birth and that changes in the
chemical composition of stellar atmospheres over time due to e.g. atomic diffusion
are sufficiently similar for all stars studied. If these assumptions are
adequate, chemical tagging
has the potential of providing interesting information on the accretion 
and star-formation history of the Galaxy. Several large on-going abundance surveys such as
APOGEE \citep{majewsky17}, {\em Gaia}-ESO \citep{gilmore12}, and GALAH \citep{bland-hawthorn16}
were designed with this purpose in mind.

A thorough discussion of the prospects of chemical tagging has been presented by
\citet{ting15}. Important parameters for the method is the dimensionality of 
chemical space, i.e. the number of abundance ratios varying independently,
and the amplitude of variations relative to the precision of the measurements.
This determines the number of chemical cells (i.e. the number of distinguishable
cloud-to-cloud variations in chemical space), which together with the number of stars
in the survey is critical for success of the method. Simulations for the GALAH
survey show that with a sample of $\sim 10^6$ stars and a number of chemical cells,
$N_{\rm cells} \simgt 5 \times 10^4$, it may be possible to detect disrupted clusters
with masses $M \simgt 10^5 \,M_{\odot}$. This assumes that the clusters are evenly
distributed over chemical space, which may not be the case. Interestingly, Ting et al.
show that an improvement of the precision ($\sigma$) of the derived abundances
is much more efficient than increasing the size of the sample,
because $N_{\rm cells} \propto \sigma ^{-N_{\rm dim}}$, where $N_{\rm dim}$ is the
dimensionality of chemical space. With $N_{\rm dim} \, = \,8$, as estimated for
the GALAH survey, a decrease of $\sigma$ by a factor of two would increase 
$N_{\rm cells}$ with a factor of 250\,!

In an application of chemical tagging on a sample of $\sim 10^5$ stars 
(including known clusters) with APOGEE abundances for 15 elements, \citet{hogg16}
were able to detect two globular clusters (M\,5 and M\,13), the Sagittarius stream,
a halo structure with high velocity dispersion, and a thin-disk structure possibly
associated with the Milky Way bar. This shows that chemical tagging can work
on high-mass structures especially at low metallicities. However,
none of the known open cluster clusters were detected. In a test of chemical tagging
of about  200 stars in 31 open clusters with abundances of 14 elements
determined from HARPS and UVES spectra,  \citet{blanco-cuaresma15} also found
that clusters could not be chemically recovered due to overlap in chemical space.
Furthermore, although \citet{mitschang14} in a chemical tagging experiment on a sample
of $\sim 700$ stars with abundances of 12 elements from \citet{bensby14} found many groups, 
they seem to represent coeval associations of stars instead of stars formed from individual 
molecular clouds implying a homogeneous chemical evolution in the Milky Way.
This suggests that it will be difficult to determine the details of
the star formation history of the disk  
from chemical tagging. The abundance precision
in the mentioned studies is on the order of 0.03\,-\,0.06\,dex. Improving the 
precisions to 0.01\,-\,0.02\,dex will enhance the potential of the method greatly
as discussed above. However, one may then run into problems with the assumption of
chemical homogeneity of stars born in a given molecular cloud related to 
dust-gas separation effects
on the composition of stars (see Sect. \ref{solartwins}), and one has to limit the sample
of stars to a small region of the \teff\,-\,\logg\ diagram to ensure that atomic diffusion
of elements modify the atmospheric composition for all stars born in a given cloud in  
the same way \citep{dotter17}.

A new way of using chemical abundances to study the chemical evolution
of stellar populations has been proposed by \citet{jofre17a}, i.e. to apply
phylogenetic techniques from biology on chemical abundances (as ``a stellar
DNA") to construct evolutionary trees. As an example, the method was used for a
sample of 22 solar twins (including the Sun), for which \citet{nissen15, nissen16}
has derived very precise differential ($\sigma \sim 0.01$\,dex) 
\xfe\ abundance ratios for 17 elements
representing the main nucleosynthesis channels, i.e. $\alpha$-capture 
(C, O, Mg, Si, S, Ca, Ti), odd-$Z$ processes (Na, Al, Sc), iron-peak processes 
(Cr, Mn, Ni, Cu, Zn), and neutron capture (Y, Ba). Based on ``chemical distances''
for pairs of stars  
an evolutionary tree with three branches was defined, but six stars could
not be assigned to any branch with statistical significance. From isochronic ages 
and kinematics of the stars, the three branches can be identified with an old (9\,-\,10\,Gyr)
thick-disk population, a relatively young (1\,-\,5\,Gyr) thin-disk population and an intermediate
population that appears discrete in several abundance ratios and may be related 
to the intermediate population suggested by \citet{fuhrmann17}. Compared to traditional
methods of detecting stellar populations from clustering in various abundance
diagrams, the phylogenetic method may add information on the sequence of stars born in
a given population. Hence, the rate of chemical evolution for a given population
can be determined by comparing
chemical distances with stellar ages. A basic assumption inherent in the method 
is that the sequence of populations in the explored region of the Galaxy has followed 
a continuous track with essentially  no separate and distinctively different
contributions, from e.g. infalling sub-systems or wide migration. 
Yet, the method seems promising and it would be
interesting to see how it works for large samples of stars having abundance
precisions on the order of 0.03\,dex, e.g. the APOGEE K giants.

\section{The Sun compared to solar twins}
\label{solartwins}

\begin{figure}
\centering
  \includegraphics[width=8.0cm]{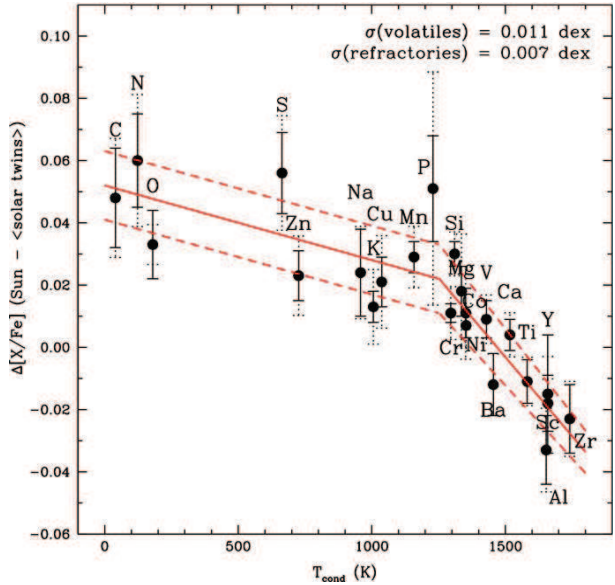}
\caption{Differences in \xfe\ between the Sun and the mean values for 11 solar twins
as a function of elemental condensation temperature. The solid red lines show a
double linear fit to the data broken at $\Tc \, = \, 1200$\,K, while the dashed lines
represent the standard deviation from the fits. The 1-sigma errors in the mean abundances
of the solar twins are shown with solid error bars, while the dotted error bars show
the 1-sigma errors of the relative abundances (including errors in both the Sun and the
twins). Reproduced from \citet{melendez09} with permission from Jorge Mel{\'e}ndez.}
\label{Fig3.Melendez}
\end{figure}

The concept of ``solar twins'' was early discussed \citep{Cayrel81, cayrel96} and used in
 defining a standard solar proxy for colour calibrations. \citet{melendez09} defined a sample
 of 11 twins, based on the Hipparcos catalogue, with maximum departures of 75 K
 in effective temperature,  of 0.10\,dex in logarithmic surface gravity and of 
  0.07\,dex in [Fe/H] from the corresponding solar values. Spectra were obtained with 
a wavelength resolution of $R=65\,000$   and a  $S/N \sim 450$ per pixel. Using a selected 
set of spectral lines and an analysis as model-independent as possible,
 abundances were derived with the individual line EW method relative to the Sun
 (as reflected in asteroid spectra). The resulting abundances for the individual stars
 were found to have intrinsic errors on the order of 0.01\,dex, and significant
 differences between the stellar average logarithmic abundance ratio [X/Fe]
  for each element X and the corresponding solar abundance were found. The difference
 have a range of about 0.08\,dex and clearly correlate with the condensation temperature
 $T_c$ for the element in a gas with solar composition, as listed by \citet{lodders03}
(see Fig. \ref{Fig3.Melendez}). \citet{bedell18}
have verified this tendency for a sample of 79 solar twins and also show that the Sun 
is not an isolated outlier but a star in the tail of a distribution of the ratio $R$ of
 abundances of ``refractories'' (high $T_c$) relative to ``volatiles'' (low $T_c$). 
On the order of $5 \%$ of the twins in the solar neighbourhood are judged to have 
ratios $R$ close to the solar ratio or less. 
An interesting example of a star which is a solar twin, also in view of its
low $R$ ratio, is M67-1194, in the old rich cluster M67 \citep{Onehag11, liu16a}.
Also our nearest solar-analogue star, $\alpha$\,Cen\,A, has a \xfe -\Tc\ trend
similar to that of the Sun \citep{morel18}.

A number of possible explanations 
for this ``Mel{\'e}ndez effect'' have been considered. The possibility that
 the stellar spectra are different from the solar one just because we observe the Sun
 at a special angle, relative to its axis of rotation, could be clearly refuted by 
solar observations at different disk locations \citep{Kiselman11}. Remaining alternatives 
included effects of Galactic evolution and nucleosynthesis, the cleansing of 
the pre-solar cloud from dust by radiation pressure from hot stars or by the Sun itself, 
the late accretion of the proto-planetary disk cleansed from refractories in dust
 by planet formation, the results of planetary engulfment to various degrees
 for different stars, and the thermal processing and outflows from the inner accretion disk.
 
It has been discussed whether the anomaly of the Sun would disappear if the
 age dependences of [X/Fe] for the solar twins were taken into account \citep{adibekyan14}, 
but \citet{bedell18} have
convincingly shown this not to be the case. The idea that the pre-solar cloud was cleansed 
by the radiation from hot stars was inspired by the possibility that the solar birth cloud 
was early on polluted by at least one  core-collapse supernova, which then implies a
 probable solar origin in a fairly rich cluster \citep{Adams10}. 
This led to a spectroscopic study of the solar-like stars in the rich and old open
 cluster Messier 67 \citep{Onehag11, onehag14}, which indeed were found to have
 abundances more close to the Sun than most solar twins, supporting the hypothesis.
 The S/N of the spectra of these apparently faint stars were, however, not comparable to those
 of the nearby twins, and further observations are warranted. Also, \citet{gustafsson18a}, 
found by model calculations that the mechanism would not be effective enough to cleanse
 most of the gas forming a full cluster from dust, and that turbulence in the cloud
 would probably strongly reduce the degree of dust-cleansing as a whole.
 Another possibility studied by \citet{gustafsson18b} is that the radiation 
from the proto-Sun itself partially cleansed the gas to be accreted in the late
 stages of the star formation process. For this mechanism to work,
 one has to assume a relatively slow accretion rate (on the order of 0.01 M$_{\odot}$/Myr),
 active during at least 1 Myr, at a stage so late that the solar convection zone
 has retracted toward the surface enough to contain only a few percent of the solar mass.

A similar constraint related to the time scale of deep convection 
(suggested by pre-main- sequence standard models to last for about 30 Myr) 
is also necessary if the cleansing of the proto-planetary disk by planet formation
 is taken as the explanation for the effect. In this case, the life-time of the gas disk,
 assumed to be finally accreted to the star, has to be prolonged considerably beyond
 the few million years usually suggested \citep[e.g.][]{Wyatt08}. Another way of
 solving this dilemma is to postulate a much more rapid retraction of the convection
 zone to the surface, due to episodic accretion onto the star, following \citet{Baraffe10}.

Yet another way to explain the Mel{\'e}ndez effect is to assume different frequencies 
of engulfed planets for different stars \citep{Melendez16, kunitomo18}, in which case the Sun 
would have had comparatively few accreted planets, or differences in the time when
 engulfment occurs, in which case the Sun would have had comparatively early such
 events when the convection zone was still deep, so that the swallowed planet(s)
 did not affect the solar surface composition very much. An attractive feature with
 this explanation is that it would explain an interesting observation by \citet{Gonzalez10},
 namely that the meteorites of chondrite type CI, which are supposed to show the 
original chemical composition of the original solar cloud, tend to have even lower 
refractory/volatile ratio than the Sun itself. The ``self-cleansing hypothesis''
 mentioned above would however, also suggest a natural explanation for this:
 the amount of accreted cleansed gas (then presumably setting the final composition
 of the disk and its chondrites) may be smaller than the mass of the convective zone of the star.

The different hypotheses concerning the origin of the Mel{\'e}ndez effect demonstrate 
how accurate studies of abundance ratios may lead to new clues to important but
 enigmatic cosmogonic processes. However, as noted in Sect. \ref{tagging},
the phenomenon also leads to some questioning
 concerning the possibility of chemical tagging as a means of detailing the 
history of Galactic evolution.

\section{Abundance differences in binaries and clusters}
\subsection{Binaries}
\label{binaries}

Since two decades abundance differences between components in binary stars have been 
searched for \citep {Gratton01}, and possible findings have been tentatively
 related to differences in the planetary systems of the components.
 These efforts have been concentrated towards wide binaries with almost identical
 components in terms of effective temperature and surface gravity,
 indeed excellent targets for differential spectroscopy in order to exploit
 the high precision achievable in a differential analysis. One good example of such
 a system is 16 Cyg A and B, both stars with similar parameters close to solar,
 the A component being 60-80 K hotter than the B component. B was found to have a planet
 more massive than Jupiter by \citet{cochran97}, while no planet has as yet been found around A.
 \citet{Gonzalez98} and \citet{laws01} found the A component to be slightly more iron rich
 than the B component (by 0.05 and 0.025\,dex, respectively) and \citet{laws01}
ascribed this difference to planet formation and even traced a tendency for the 
abundance difference with condensation temperature. \citet{ramirez11} also found
 the A component to be more metal-rich than the B component, by 0.04\,dex$ \pm 0.01$\,dex,
 but did not trace any significant difference in this respect between refractories and volatiles.
 \citet{tuccimaia14}, however, concluded that the volatiles depart by 0.03\,dex at a mean in
 between the components and found the refractories to show differences that grow with
 condensation temperature for the element. This was
confirmed by \citet{nissen17}. The total amplitude in the variation $\Delta$[X/H] between
 the components is 0.02\,dex. The authors estimate the core of the rocky planet,
 which is supposed to bind the refractories that would else have ended in the
 convection zone of the B component, to be of the same order of magnitude as the
core of Jupiter.

\begin{figure}
\centering
  \includegraphics[width=11.0cm]{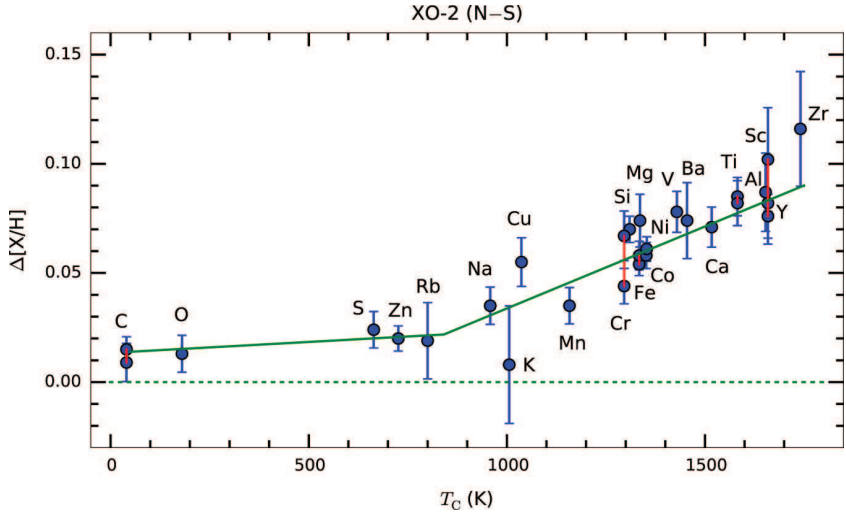}
\caption{The abundance difference between the binary stars XO-2N and XO-2S as 
a function of the elements' condensation temperature. The solid green line is a double
linear fit to the data broken at $\Tc \, = \, 840$\,K. Red vertical lines connect 
\xh -values derived from two species of the same element, e.g. \CrI\ and \CrII .
Reproduced from \citet{ramirez15} with permission from Ivan Ram{\'{\i}}rez.}
\label{Fig5.Ramirez}
\end{figure}

Another interesting example is the binary XO-2, consisting of two very similar K dwarfs 
with estimated effective temperatures around 5300 K and masses of about 0.97 solar masses
 \citep{biazzo15, ramirez15}. Both stars have planetary systems. In this case,
 the amplitude in $\Delta$[X/H] extends over 0.10\,dex (see Fig. \ref{Fig5.Ramirez}),
 which is much since the low
 effective temperatures indicate that the mass of the convective zone is much greater
 than higher up along the main sequence.  Therefore, a difference of about 50 Earth masses
 of refractory material in between the stars is needed to affect so much mass
 sufficiently and explain the differences. It should be noted that the late spectral
 type of these stars makes the abundance analysis more difficult due to the rich
 and blended spectra \citep{teske15}. Internal errors in the abundance differences
 between the stars vary between 0.02\,dex for some elements up to 0.05\,dex or even more.

Yet another interesting binary is WASP-94 A and B. These dwarf stars are more metal rich
 and hotter than the Sun, of spectral types F8 and F9, and both stars host a hot 
Jupiter-like planet. According to \citet{teske16a} abundance differences between the
 A and B components show a unique pattern  compared  with those of the other binaries
 known to show different abundances: A is {\it poorer} in volatiles than B by 0.02\,dex
 {\it but richer} than B in refractories by 0.01\,dex. Although these effects are small,
 they are still significant, at least if the internal precision in the abundance differences
 claimed by the authors to be about 0.005\,dex is adopted as realistic errors.

The two components in the binary HAT-P-4 were found by  \citet{saffe17} to be 
somewhat hotter (by +250 K) and more metal-rich than the Sun (by up to 0.3\,dex).
 The A component has a hot Jupiter in orbit while no planet has been found around B. 
In a detailed line-by-line differential analysis of spectra with high S/N,
 \citet{saffe17} found the A component
to be significantly richer in refractory elements than B by as much as about 0.1\,dex
 (one of the greatest internal differences as yet found among binary abundances),
 while the stars have
more similar abundances in volatiles. Another noteworthy effect is that A has a
 significantly higher Li abundance, contrary to the suggestion 
by \citet{delgadomena14} that close planets cause
more efficient Li depletion.   A scenario involving formation and later infall
 of rocky planets and planetesimals onto the A component,
 while the giant planet spirals inwards, is proposed by the authors.

Another quite interesting wide binary is the two co-moving solar-like stars HD240430 
and HD240429, explored by \citet{oh18}.
The authors find, again in a study at high resolution and a S/N of about 200,
 that the HD240430 has roughly solar-like volatile abundances while its refractories
are enhanced by about 0.2\,dex with
respect to the Sun and its companion (a true record holder in this respect).
  HD 240429 has both volatiles and refractories close to solar. We also note that HD240430
has a significantly higher Li abundance, and that none of the stars is known to have a planet today.

One should not be misled by the selection of examples above to believe that all
 solar-type binary stars show pronounced abundance differences
scaling with condensation temperatures. One example of the contrary is the wide
 system HD80606/ 80607 where the two components were
found in a high-precision analysis by \citet{ Liu18} to depart in metallicity 
by 0.013 $\pm 0.002$\,dex (!) but with no clear variation with condensation
temperature.  

The results for binary compositions clearly demonstrate the possibilities of 
strictly differential abundance analysis to trace interesting effects of
 fractionation of chemical elements in the
early evolution of stars and planetary systems. They again suggest that one should
 not overemphasize the possibilities of chemical tagging in the study of Galactic evolution.
 A useful summary of the results achieved in abundances studies for wide binaries 
is provided in an appendix by \citet{oh18}.

\subsection{Clusters}
\label{clusters}

Since long, the chemical homogeneity of star clusters has been discussed.
 Chemical differences have been found 
among stars in the same cluster and may be consequences of stellar
 evolution processes, such as
dredge-up of processed matter due to CNO burning \citep{kraft94},
 or destruction of Li \citep{Boesgaard86} and Be \citep{Smiljanic10}, 
or of ``diffusion''  -- or sooner the interaction of radiative pressure,
 turbulent and other mixing, and sinking in the gravitational field
  \citep{korn07, Nordlander12, Gruyters14, Bertelli18}.
It seems clear that the effects found are not to be referred to departures
from LTE; differences in atmospheric structure should, however, be further explored.
 The study of these processes in clusters involves 
comparison of stars in different evolutionary stages,
 i.e. with different parameters which, although the results of the studies
 may be truly interesting, makes it difficult to achieve the degree of
 accuracy discussed here. In the diffusion studies, this technical difficulty may
 be partly circumvented by comparison of stars with very similar
 effective temperatures, although on different branches
in the colour-magnitude diagram, such as the main-sequence and
 subgiant-giant branches, where the effects of diffusion are
 expected to be different. Of a special interest is, however,
inter-comparison between very similar stars -- abundance differences between
 these could indicate inhomogeneities in the gas cloud in which the stars were formed.
 Such differences would be contrary to the expectation that the proto-cluster
 would be homogenized before the first supernova \citep{Krumholz09}  and have
 important bearings on the cluster- and star-formation processes. Also,
 such findings would give fundamental constraints on the possibilities of
 studying Galactic evolution by ``chemical tagging''. 

For the globular clusters, star-to-star variations with increased He, N and Na,
and correspondingly decreased C and O abundances have been known and studied since long.
Abundance variations of C and N for stars of similar luminosity on
 the giant branch were traced by \citet{Bell80} and clear evidence of a
Na-O anti-correlation was found by \cite{kraft93} for giants in M\,13
and by \citet{gratton01b} for turnoff stars in NGC\,6752. The effects were interpreted
 as signatures of hot H burning and effects of multiple generations of stars.
\citet{Carretta09} published homogeneous O, Na, Mg , Al and Si abundances
based on high-resolution spectra for more than 200 red giants in 17 Galactic
 globular clusters with widely different over-all metallicities. These authors found
 star-to-star variations in O and Na abundances greater than the expected errors
 (the latter judged to be typically 0.05 to 0.08\,dex in [Na,Mg,Al,Si/Fe]),
 in all clusters. Significant variations in Al were found for all but a few clusters,
 while variations in Mg and Si were found for just a handful of them.
 The situation was recently reviewed by \citet{Bastian17} who demonstrate its
 complexity in terms of different abundances patterns as well as other relevant
 cluster properties such as split main-sequences.

In order to investigate further
what determines the existence of multiple stellar generations and other factors of
possible importance, high precision spectroscopic abundances are important
as shown by \citet{yong13} in a study of 38 red giant stars in NGC\,6752.
This is one of the globular clusters for which large star-to-star abundance
variations of C, N, O, Na, Mg, and Al have been detected, but
which was believed to have no abundance variations for heavier elements.
However, in a line-to-line differential analysis of high quality VLT/UVES spectra
providing differential abundances with precisions of 0.01 to 0.05\,dex,
Yong et al. found that many of the heavier elements, e.g. Si, Ca, Ti, Y, and Eu,
even seem to have abundance variations that are typically  a factor of 2 to 3 larger than
the measurement errors and positively correlated with the Na variations.
The authors demonstrated that the differential effects found were not a result 
of differences in effective temperatures of the stars intercompared, nor
the result of their choice of reference stars.
The resulting abundance variations 
could be due to variations in the He abundance, because for a fixed
mass fraction of heavy elements an increase of the helium mass fraction ($Y$)
causes a corresponding decrease of the hydrogen mass fraction  and hence
an increase of the abundance of all heavy elements relative to hydrogen.   
The helium abundance  variations in NGC\,6752, $\Delta Y = 0.03$ as estimated
from colour-magnitude diagrams \citep{milone13}  explain, however,
only half of the heavy element abundance variations, so in addition there
may be inhomogeneous chemical evolution effects in the proto-cluster gas.
Obviously, such high precision studies of other globular clusters would
be interesting.

A clear demonstration of abundance scatter in an open cluster beyond the effects caused
 by stellar evolution discussed above was presented by
 \citet{liu16b} for the Hyades cluster; 
see also this paper for references concerning previous unsucessful attempts in this direction.
 The authors carried out differential line-by-line analyses of 19 elements in
 16 solar-type stars in the cluster,
using high-resolution spectra with S/N $\sim 350-400$.
 The intrinsic scatter for an element 
in between the Hyades stars (0.02 to 0.04\,dex) is significantly greater 
than the average measurement errors (0.01 to 0.02\,dex)
for several of the elements, including Si, Ca, Fe and Ni, 
and the different deviations from the mean for all cluster stars 
for these elements also correlate. The differences show no 
significant variation with
the condensation temperatures of the elements and are found not to be
 the result of diffusion. Among the possible explanations is 
an original inhomogeneous composition of the
proto-cluster, the explosion and ejection of an early supernova in the
 cluster-forming cloud, or the pollution by an in-coming gas cloud
 with different chemical composition. 

Another example of abundance differences among members of an open cluster
was recently published by \citet{lorenzospina18}. For five solar-type stars in the relatively
young (age $\sim 100$\,Myr) Pleiades cluster, UVES spectra with a resolving power of 
$R = 75\,000$ and $S/N \sim 400$ were obtained and used to determine differential 
abundances of 19 elements with precisions on the order of 0.02\,dex.
The star-to-star abundance differences are clearly correlated with elemental
condensation temperature; there are no significant differences for the volatile elements,
but for refractory elements with \Tc\ in the range from 1300\,K to 1700\,K, the differences
(relative to the mean for the Pleiades cluster) range from $-0.06$ to +0.04\,dex.
The authors suggest that this may be explained by  
engulfment of rocky planets, i.e. the star with the most negative \Tc\ trend
has a stable planetary system (like the Sun), whereas the other stars have 
accreted different amounts of rocky planetary material leading to an increase of the  
refractory-to-volatile ratio.

These first results for two of our most near-by open clusters certainly suggest 
that chemical analyses at similar precision of stars in additional clusters would be interesting.

\section{Abundance-age relations for solar twins}

\subsection{Constraints on nucleosynthesis}
\label{nucleo}

Models of nucleosynthesis and of Galactic chemical evolution are
often tested by comparing predicted and observed $\xfe - \feh$
relations under the assumption that \feh\
increases smoothly with time at a given position in the Galaxy.
For stars in the solar neighbourhood, there is however, a spread in \feh\ of more
than 0.3\,dex at a given age \citep{edvardsson93, nordstrom04}, which may
be due to spurious infall of metal-poor gas or migration of stars in the disk. Therefore,
abundance ratios as a function of stellar {\em age} provide a more direct test
of chemical evolution models. Ages of sufficiently high precision for this purpose
may be obtained for solar twins, because the deviation of a star from the Sun in a
$\logg - \teff$ or $L - \teff$ diagram may be converted to a difference in age 
via isochrones.  

In the following, we will discuss relations between abundance ratios and age
for solar twins in the solar neighbourhood obtained from 1D, LTE model atmosphere 
analyses of HARPS spectra with $R = 115\,000$ and $S/N > 500$ by
\citet{nissen15, nissen16} and \citet{nissen17} for 23 stars and by
\citet{spina18} and \citet{bedell18} for 79 stars. 
The total range in \feh\ for these samples goes from $-0.15$\, to +0.13\,dex,
but the large majority of stars have \feh\ within the $\pm 0.10$ range used
in the definition of a solar twin.
\citet{nissen15} included non-LTE corrections for some elements, but for solar
twins the differential corrections with respect to the Sun are almost negligible. 
The high precision obtained in these studies is testified by an excellent
agreement for 14 stars in common between the two sets of works. The 
mean differences and standard deviations (Spina \& Bedell -- Nissen) are:
$\Delta  \teff = 0 \pm 6$\,K, $\Delta \logg = -0.003 \pm 0.014$,
$\Delta \feh = -0.006 \pm 0.005$, and $\Delta {\rm age} = 0.4 \pm 0.5$\,Gyr\,\footnote{
The age comparison refers to ages derived from Yonsei-Yale isochrones
\citep{demarque04}; as shown by \citet{nissen16}, ASTEC isochrones
\citep{jcd08} lead to a somewhat different age scale with 10\% higher ages
for the oldest ($\sim 10$\,Gyr) stars, probably because the ASTEC models include 
heavy element diffusion in contrast to the Yonsei-Yale models.}.
For the abundances, the standard deviations of the comparison range from
0.006\,dex for the best determined elements (Na, Si, Ca, Cr, Ni) to about
0.010\,dex for C, Mg, Al, S, Sc, Ti, Mn, Cu, Zn, Y, and Ba. For oxygen the scatter
is higher, i.e. 0.025\,dex, because of relative large errors of the EW of
the weak forbidden \OI\ line at 6300\,\AA\ used by \citet{nissen15} to derive O abundances.
\citet{bedell18} used instead the 
\OI\ triplet at 7774\,\AA\ in Magellan Telescope MIKE spectra of lower quality than
the HARPS spectra applied for the other elements.

\begin{figure}
\centering
\includegraphics[width=12cm]{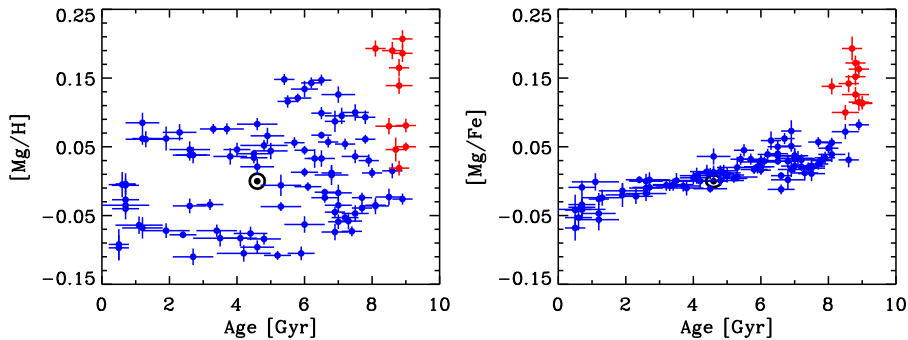}
\caption{\mgh\ and \mgfe\ of solar twins as a function of stellar age.
High-alpha stars ($\mgfe \geq 0.1$) are shown in red color; the others in blue.
Based on data from \citet{nissen15}, \citet{nissen17}, and Bedell et al. (2018).}
\label{fig:mgh.mgfe}
\end{figure}

In the works cited above, age trends of abundance ratios with respect to iron, 
i.e. \xfe , have been studied. There are, however, two sources of iron:
Type II SNe contribute on a time scale of $\sim 10$\,Myr 
while Type Ia SNe on a scale of 1\,-\,1.5\,Gyr \citep{matteucci09}.
This complicates the interpretation of 
\xfe -age relations. We will instead consider trends of \xmg\ with age.
The main isotope of Mg, $^{24}$Mg, is made by carbon and neon burning in massive stars
and dispersed ``instantaneously'' to the interstellar medium by Type II SNe. 
Hence, the \xmg -age relations may be used to investigate if other elements are also made 
in high-mass stars or if lower mass stars (including Type Ia SNe) make a delayed contribution. 

As seen from Fig. \ref{fig:mgh.mgfe}, \mgh\ of the solar twins has a total spread  of about
0.2\,dex at all ages. In contrast, there is a remarkable tight relation between
\mgfe\ and age, except for the old 8-9\,Gyr stars, among which we have marked 
high-Mg stars ($\mgfe > 0.1$) in red. There is also a hint of a cosmic scatter in \mgfe\ for
the youngest stars and in the age bin 6-7\,Gyr, but from 2 to 6\,Gyr the rms scatter of \mgfe\
at a given age is only 0.01\,dex, i.e. of the same order of size as
the estimated precision of the derived abundances.

The dispersion in the metallicity-age relation for stars in the solar neighbourhood
is often explained as due to orbital mixing (migration) of stars born at different galactocentric
distances, $R_G$, in a Galactic disk with a radial abundance gradient \citep[e.g.][]{minchev14}. 
For a gradient of $\delta \mgh / \delta R_G = -0.05$\,dex per kpc, migration
should be effective over a Galactic region corresponding to 
$R_G$ from 6 to 10\,kpc in order to explain the range in \mgh\ for solar twins in
the solar neighbourhood. On the other hand, the Galactic gradient in \mgfe\ should be
vanishingly small in the age range 2 to 6\,Gyr to explain the low dispersion in \mgfe .
This requires that the ratio of Type II and Type Ia SNe has 
evolved in the same way for $R_G$  between 6 and 10\,kpc. Alternatively, as suggested by
\citet{edvardsson93}, the dispersion in \mgh\ could be due to infall of metal-poor gas
triggering star formation. 

\begin{figure}
\centering
\includegraphics[width=12cm]{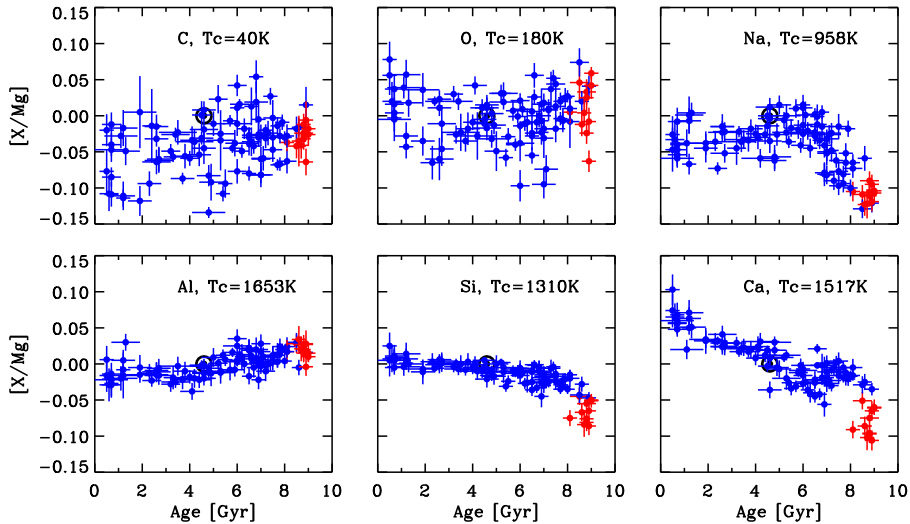}
\caption{Abundances of C, O, Na, Al, Si, and Ca relative to Mg as a function of stellar age
for solar twins.
High-Mg stars are shown in red like in Fig. \ref{fig:mgh.mgfe}. \Tc\ is 
the elemental condensation temperature in a solar-composition gas \citep{lodders03}.
Based on data from \citet{nissen15}, \citet{nissen17}, and Bedell et al. (2018).}
\label{fig:CONaAlSiCa.Mg-age}
\end{figure}
\begin{figure}
\centering
\includegraphics[width=12cm]{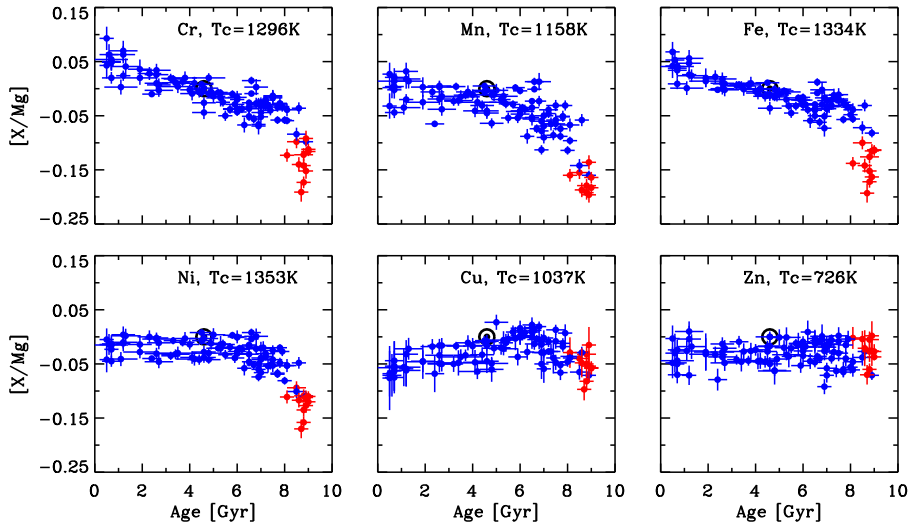}
\caption{Abundances of Cr, Mn, Fe, Ni, Cu, and Zn relative to Mg as a function of stellar age
for solar twins. 
Based on data from \citet{nissen15, nissen16}, \citet{nissen17}, and Bedell et al. (2018).}
\label{fig:CrMnFeNiCuZn-age}
\end{figure}

The \xmg -age relations for some key elements are shown in Figs. \ref{fig:CONaAlSiCa.Mg-age}
and \ref{fig:CrMnFeNiCuZn-age}. As seen, the scatter at a given age tends
to be higher than the estimated errors bars for elements having a condensation
temperature significantly lower than that of Mg (\Tc = 1336\,K), i.e. C, O, Na, Mn, Cu, and Zn.
This may be due to variations in the \xh -\Tc\ trends
related to dust-gas separation in star-forming gas clouds.
Interestingly, the Sun lies close to the upper envelope of the \xmg -age distribution
for the low-\Tc\ elements, showing (as also discussed in Sect. \ref{solartwins}) that the Sun has
a high ratio between volatile and refractory elements compared to solar twins with an
age around 4.6\,Gyr.

The high dispersion in \cmg\ and \omg\ makes it difficult to see if there is an  
overall trend with age but it does not seem to be the case
suggesting that C and O are primarily made in massive stars like Mg.
In the case of oxygen, this is also predicted from nucleosynthesis calculations,
but from chemical evolution modelling of observed 
\co -\oh\ and \cfe -\feh\ relations, it has been 
suggested that low-mass stars have provided up to half of the carbon in the
Galactic disk \citep{chiappini03, carigi05}. If so, there should have been a rise
of \cmg\ towards younger ages, but if anything, \cmg\ is declining slightly.   
We conclude that low-mass stars are not contributing significantly to the carbon
abundance of solar metallicity disk stars in agreement with the conclusions
of \citet{gustafsson99} based on  a study of the \co -\feh\ relation for dwarf stars in
the solar neighbourhood.
 
As seen from Fig. \ref{fig:CONaAlSiCa.Mg-age}, \namg\ rises steeply with time
(i.e. decreasing age)
for stars older than about 6\,Gyr. Sodium is made by hydrostatic carbon burning in massive stars
with a yield that increases with the neutron-excess, which depends on the C and O abundances of
the Na-producing stars \citep[e.g.][]{kobayashi06}. Therefore, the rise of \namg\ suggests
that the chemical evolution in the early Galactic disk was dominated by massive stars
with sub-solar abundances of C and O. For stars younger than about 6\,Gyr, \namg\
is nearly constant indicating that the nucleosynthesis of Na has mainly occurred in solar-metallicity
massive stars. Sodium production in the hydrogen burning shell of low and intermediate mass AGB stars
via the Ne-Na cycle \citep{karakas10} does not seem important.

The near-constant trend of \almg\ with stellar age supports that aluminium is made
in massive stars by carbon and neon burning along with magnesium.
\simg\ shows a slight rise with time, which is probably due to a non-negligible
contribution from Type Ia SNe relative to the contribution from Type II SNe
\citep[e.g.][]{kobayashi06}. Among the other $\alpha$-capture elements, Ti (not shown) behaves
like Si, but Ca is a surprise; as seen from Fig. \ref{fig:CONaAlSiCa.Mg-age}, 
\camg\ increases by about 0.15\,dex from the oldest to the youngest stars,
i.e. similar to the increase of \femg . Excluding the high-Mg stars (red points),
\cafe\ is constant as a function of age with a scatter less than 0.01\,dex.
Thus, the contribution of Type Ia SNe to calcium seems as important as 
their contribution to iron, which is not predicted by nucleosynthesis
calculations \citep[e.g.][]{kobayashi06}. Alternatively, there could be
an additional source of Ca. An intriguing possibility is low-luminosity
SNe with presumable low initial masses like SN\,2005E for which the yield of 
Ca is 5 to 10 times higher than that of classical Type Ia SNe \citep{perets10}.
\citet{frohmaier18} estimated the volumetric rate of these
so-called Ca-rich transients in the local Universe to be about 50\% of the Type Ia SNe rate,
which can explain the enhanced Ca/Fe ratio in the intracluster
medium of galaxy clusters.

The age trends for some of the iron-peak elements are shown in Fig. \ref{fig:CrMnFeNiCuZn-age}.
\crmg\ has nearly the same dependence on age as \femg\
indicating that the ratio of contributions from Type Ia and Type II SNe is about 
the same for Cr and Fe. \mnmg\ and \nimg , on the other hand, show a more flat
distribution for ages  below 6\,Gyr, which is a surprise, because nucleosynthesis
calculations predict that the relative contribution of Mn and Ni from Type Ia SNe is more important
than in the case of Fe \citep[e.g.][]{kobayashi06}. \znmg\
is nearly constant over the whole age range, suggesting that Zn is made 
in massive stars along with Mg,
whereas \cumg\ shows a more complicated behaviour, which may reflect that
there are several possible ways of Cu production 
including the metallicity-dependent weak $s$-process in massive stars
\citep{pignatari10} and explosive nucleosynthesis in Type II SNe  
\citep[see discussion in][]{romano07}.

\begin{figure}
\centering
\includegraphics[width=12cm]{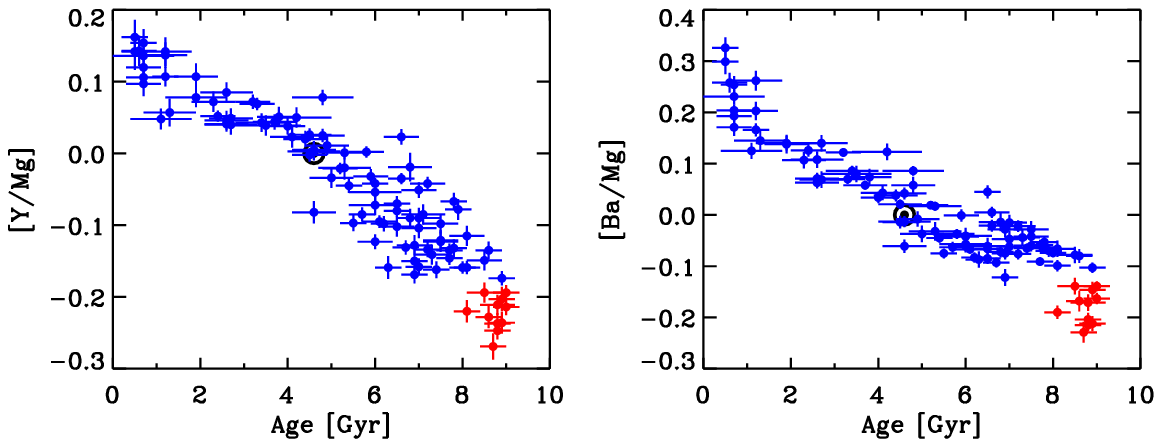}
\caption{Abundances of Y and Ba relative to Mg as a function of stellar age
for solar twins.
Based on data from \citet{nissen15, nissen16}, \citet{nissen17}, and \citet{spina18}.}
\label{fig:YBa.Mg-age}
\end{figure}

The age trends of the neutron capture elements Y and Ba 
are shown in Fig. \ref{fig:YBa.Mg-age}.
Yttrium represents abundances of elements belonging to the first peak of
the $s$-process (the {\em light} $s$-process elements, Sr, Y, and Zr) and barium
the second peak (the {\em heavy} $s$-process elements, Ba, La, and Ce).
As seen both \ymg\ and \bamg\ show a large increase with
decreasing age. This rise with time can be ascribed to the importance of the $s$-process in
low-mass AGB stars \citep{busso01, maiorca12, karakas16}. As time goes, stars with 
lower and lower mass contribute to the production of Y and Ba. There are, however, differences in the
evolution of Y and Ba. \ymg\ shows the steepest rise with time for the oldest
stars, whereas \bamg\ is rising more steeply for the youngest stars.
As shown by \citet[][Fig.8]{spina18}, \bay\ decreases from 9 to 6\,Gyr, reaches
a minimum around the solar age, and increases with time for the youngest stars.
Qualitatively, this behaviour can be explained if the
$s$-process elements in the oldest stars come from metal-poor AGB stars with
a high ratio between {\em heavy} and {\em light} $s$-process elements \citep{busso01}, whereas
the evolution at younger ages is dominated by solar-metallicity AGB stars.
According to \citet[][Fig. 15]{karakas16}, the heavy-to-light $s$-process 
ratio increases by $\sim 0.3$\,dex, when the mass of the AGB models decreases from
3 to 1.5 $M_{\odot}$.

As seen from Fig. \ref{fig:YBa.Mg-age}, \bamg\ has a significant scatter
($\sim 0.2$\,dex) among the youngest solar twins (age $< 1.5$\,Gyr). Furthermore, very large 
barium abundances, $\bafe \simeq 0.6$, have been found for F and G dwarfs
in young open clusters \citep{dorazi09}, whereas stars in young associations have
lower values, $\bafe \sim 0.25$, \citep{dorazi12, reddy15}. According
to \citet{reddy17}, this large spread in the derived \bafe\ values for young stars may be 
spurious due to problems in determining Ba abundances from the rather strong \BaII\
lines available. In a precise study of 24 solar twins with Ba abundances 
derived from the \BaII\ 5853\,\AA\ line, they find a 
correlation between \bafe\ and chromospheric activity. They suggest
that the derived Ba abundances of the young active stars are too
high, because the microturbulence in the upper atmosphere is underestimated in a classical 
1D model-atmosphere analysis. Their proposal is supported by the fact that the 
abundances of other second-peak $s$-process elements, La and Ce, derived from weak lines
do not show the same high overabundances as Ba. In general, abundances
derived from strong lines formed in the upper part of stellar atmospheres may be
systematically in error if analysed with 1D atmospheres adopting a depth-independent
microturbulence parameter. 3D model atmospheres may alleviate this problem, but abundances
based on weak lines are to be preferred. Unfortunately, no such lines are available for Ba. 

Based on a very tight, linear correlation between \ymg\ and stellar age for 21 solar
twins, \citet{nissen15} proposed that the Y/Mg ratio can be used to determine
precise ages. According to Fig. \ref{fig:YBa.Mg-age} based on many more
twins, there seems however to be a significant cosmic scatter in the \ymg -age relation
corresponding to $\sim 1$\,Gyr in age. Furthermore, \citet{feltzing17}
have shown that \ymg\ decreases with \feh\ at a given age, which makes sense, 
because at low metallicity there are fewer Fe atoms per neutron to act as seed
for the $s$-process. Thus, the use of \ymg\ or other abundance ratios
as chemical clocks would require calibrations including a metallicity term.
It should also be emphasized that the relations shown in Fig. \ref{fig:YBa.Mg-age}
refer to stars in the solar neighbourhood and may be different at other Galactocentric
distances or in other galaxies with a different chemical evolution history. 
Thus, \citet{vanderswaelmen13} find
values of \ymg\ and \bamg\ for solar metallicity stars in the Bar of the 
Large Magellanic Cloud that are significantly higher that the values for 
solar twins at any age.

The \xmg -age relations shown in Figs. \ref{fig:CONaAlSiCa.Mg-age}, \ref{fig:CrMnFeNiCuZn-age},
and \ref{fig:YBa.Mg-age} are only valid for
solar twin stars, i.e. for a narrow range in
metallicity, $-0.1 \simlt \feh \simlt 0.1$. At other metallicities, the relations may be different
especially for elements with metallicity-dependent yields such as Na and Y.
This means that more information on nucleosynthesis and chemical evolution could
be obtained if relations between differential abundance ratios and age of the same quality as those
for the solar twins were available at other metallicities. 

Atomic diffusion of elements is another problem to be considered if
abundance-age relations are used as a constraint on chemical evolution models.
According to \citet{turcotte98}, the combined effect of
gravitational settling and radiative acceleration has decreased the initial abundance 
of heavy elements in the atmosphere of an 8\,Gyr solar twin star by  
$\sim 12$\%. The amount differs by up to 2\% between the elements; hence, the
effect on the amplitude of \xmg -age relations shown in  Figs. \ref{fig:CONaAlSiCa.Mg-age} 
and \ref{fig:CrMnFeNiCuZn-age} is less than 0.01\,dex. If the stars studied cover a larger
range in \teff , differential diffusion effects may be more important.
In this connection, we note that diffusion is not affecting the atmospheric
composition of K giants  due to their deep convection zones. Therefore, it would be
interesting to obtain precise differential abundances of K giants in clusters with 
a range of ages in order to test the \xfe - age relations obtained for solar twins.

Finally, a few words about lithium. Its abundance, 
$A {\rm (Li)} = {\rm log}\,(N_{\rm Li} / N_{\rm H}) +12$,
in the atmospheres of FGK dwarfs and subgiants
depends on stellar mass and age due to proton-capture destruction of Li
at the bottom of the convection zone and on stellar
age because of Galactic evolution of Li \citep[see e.g.][]{prantzos17}. 
For solar twins with masses close to $1 \, M_{\odot}$, 
precise lithium abundances have been derived from the 
the \LiI\ 6708\,\AA\ resonance line in high S/N HARPS spectra, 
but conclusions from different works do not agree.
\citet{delgadomena14} find no significant correlation of $A {\rm (Li)}$
with age for stars older than 
$\sim 2$\,Gyr but a lower average Li abundance of planet hosting stars than 
the Li abundance of stars without
detected planets, which they ascribe to additional rotationally induced mixing at
the bottom of the convection zone in stars with planets. \citet{carlos16}, on the other
hand, find a declining trend of Li abundance with age for the majority
of solar twins (including the Sun and a few stars with detected planets), i.e.
from $A {\rm (Li)} \simeq 2.2$ at 1\,Gyr to $A {\rm (Li)} \simeq 0.6$\,dex at 8\,Gyr.
The trend corresponds reasonable well to predictions of Li depletion in 
$1 M_{\odot}$ stellar models with non-standard mixing calibrated on the Sun.

\subsection{Constraints on exoplanet compositions}
\label{exoplanets}

The relations between abundance ratios and age for solar twins may also
be used as constraints on the composition of planets around
stars. In addition to mass and radius, the proportions of C, O, and
rock-forming elements like Mg, Si, and Fe determines the structure, mineralogy,
and geodynamics of terrestrial planets and therefore should be taken
into account when discussing the potential for habitability
\citep[e.g.][]{bond10,carter-bond12, unterborn14, dorn15, thiabaud15}.
Of particular importance are the C/O number ratio.
The Sun has  $\conum = 0.55 \pm 0.08$ \citep{asplund09}, but if
a host star and hence the proto-planetary disk has
$\conum \simgt 0.8$ there is a possibility for the existence of ``Carbon'' planets 
containing large amounts of graphite and carbides instead of Earth-like silicates
\citep{kuchner05, bond10}. The Mg/Si ratio is also important; the Sun has
$\mgsinum = 1.20 \pm 0.14$ \citep{scott15}, but if the protoplanetary disk
has Mg/Si less than 1.0, terrestrial planets will have a Mg-depleted mineralogy
different from that of the Earth \citep{carter-bond12}.

\begin{figure}
\centering
\includegraphics[width=12cm]{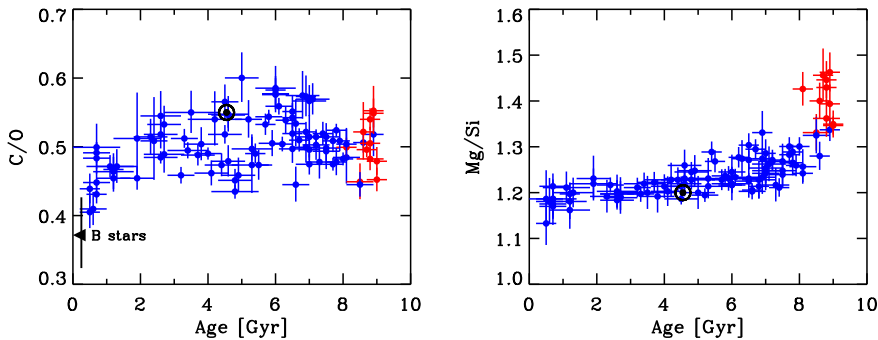}
\caption{The C/O and Mg/Si number ratios in solar twins as a function of stellar age.
Error bars refer to values  relative to the solar ratios
$\conum = 0.55 \pm 0.08$ \citep{asplund09} and $\mgsinum = 1.20 \pm 0.14$
\citep{scott15}. In the left panel, the C/O ratio in young B-type stars
in the solar neighbourhood \citep{nieva12} is marked.
Based on data from \citet{nissen15}, \citet{nissen17}, and \citet{bedell18}.}
\label{fig:CO.ratio}
\end{figure}

Determinations of C/O ratios by \citet{delgadomena10} and
\citet{petigura11} suggested that a substantial fraction of solar-type stars
have $\conum > 0.8$, but later more precise work did not confirm this
\citep{nissen13, nissen14, brewer16b, suarez-andres18}; C/O rises with
\feh\ but stays below 0.8 even at the highest metallicities, $\feh \sim +0.4$.
For solar twins, C/O is also well below 0.8 as seen 
from Fig. \ref{fig:CO.ratio}. There is some indication of a cosmic scatter in C/O
at a given age and a decline of C/O towards young ages. This trend agrees with the
low C/O ratio ($\conum = 0.37 \pm 0.05$) for young B-type stars in the solar
neighbourhood determined by \citet{nieva12} from a careful non-LTE study
of high-S/N spectra. They interpreted the low C/O of B stars relative to the 
solar ratio in terms of a strong radial gradient of C/O in the Galactic disk
combined with migration of the Sun from an assumed birthplace at a distance of 
5-6\,kpc from the Galactic centre. The data for solar twins suggest, however,
that the difference in C/O may instead be due to the difference in age between the Sun and 
young B stars.

As the Mg and Si abundances in solar-type stars are close to each other, it requires
high-precision data to determine if Mg/Si in a given star is above or below 1.0. 
In a 1D, LTE abundance study
of 849 F, G, and K dwarfs in the solar neighbourhood based on high-S/N KECK/HIRES spectra,
\citet{brewer16b} found a Mg/Si  distribution significantly broader than the
estimated error distribution  and concluded that about 40\% of the stars have 
$\mgsinum < 1.0$. \citet{suarez-andres18} also found $\mgsinum < 1.0$ values
in an analysis of 499 solar-type stars with high-S/N HARPS spectra; 15\% of the
stars harbouring high-mass planets have $\mgsinum < 1.0$ and Mg/Si decreases with 
increasing \feh\ reaching $\sim 1.0$ at $\feh = +0.3$. On the other hand, 
it is seen from  Fig. \ref{fig:CO.ratio} that
the Mg/Si ratio in solar twins is well above 1.0 when adopting
a reference solar ratio of $\mgsinum = 1.20 \pm 0.14$ as derived by \citet{scott15}
from an analysis of photospheric lines based on 1D non-LTE corrections
to 3D LTE synthetic spectra.
If instead the meteoritic ratio of CI carbonaceous chondrites,
$\mgsinum = 1.05 \pm 0.05$ \citep{lodders03}, is adopted for the solar reference value
as in \citet{brewer16b} and \citet{suarez-andres18}, the Mg/Si ratio of solar twins
approaches 1.0 for the youngest stars but only one of these stars has $\mgsinum < 1.0$.

We conclude that the solar twin data provide no evidence for the existence of
stars with $\conum > 0.8$ or $\mgsinum < 1.0$, which suggests
that terrestrial planets around solar metallicity stars have compositions similar 
to that of the Earth. The more metal-rich stars,  those with $\feh \sim +0.3$,
are more likely to have terrestrial planets with a composition and structure different from
that of the Earth, but detailed 3D, non-LTE analyses are needed to improve the accuracy
of the derived abundances for the corresponding stars.
Further 3D, non-LTE studies of the Mg/Si in the Sun would also
be interesting given that the photospheric ratio is higher than the ratio in
meteorites.

Considering that several terrestrial planets have been discovered in orbits
around M dwarfs, e.g.  one planet around Proxima Centauri \citep{anglada16} and 
seven planets around TRAPPIST-1 \citep{gillon17}, and that many more are 
expected to be discovered during the TESS and Plato missions, there is
a high interest in determining precise parameters of M dwarfs.  
In particular, it would be interesting to determine ages of M dwarfs
from abundance ratios that are closely correlated with age due to
Galactic chemical evolution (see Sect. \ref{nucleo}). As discussed in 
Sect. \ref{coolstars}, \alphafe\ may be determined with 
internal precisions of $\sim 0.05$\,dex from near-infrared spectra of   
M dwarfs. It remains, however, to be seen if this precision can also
be obtained for elemental ratios that are sensitive to age, 
such as $s$-process elements relative to $\alpha$-capture elements, e.g. Y/Mg.

\section{Conclusions and recommendations}
\label{conclusions}

The quantitative chemical analysis of stars has developed in a very impressive way. 
With pioneering work in the 1920'ies and 1930'es by \citet{Payne28},
\citet{Russell29}, \citet{Unsold38}, \citet{Stromgren40},
 and others, it provided some important basic arguments for the famous B$^2$FH article
 \citep{burbidge57} on the synthesis of elements in stars.
Yet, in the 1960'ies the results of the analyses were often found to be
in error by more than a factor of two, due to systematic errors which
 were difficult to judge. This made it even possible,
as late as in the early 1970'ies \citep{Worrall72}, to question whether the abundances
 should at all be taken seriously.  The rather low quality of abundance data was
 realized by many of the
spectroscopists active then; even if they in some cases found rather high
internal precision in their studies, the possible systematic errors often raised 
question marks, even among optimists.

Since then, however, the situation has improved, and, as in other areas of science,
 this is to a great extent due to improvements in technology. Larger telescopes,
 more efficient spectrometers,
detectors with linear response, new algorithms and computers for ever better and more detailed
 modeling and data handling, and great improvements in atomic and molecular data,
 in all these respects the situation is now totally changed. 
 It is also important to realize the fundamental contribution by a number
 of individuals, often specialists with limited recognition outside their own speciality areas,
 who made the progress possible by devoted work on 
 spectral line identifications, transition rates, non-LTE radiative transfer,
 turbulent convection, spectrometer stability, detector development, data-base handling, etc.

Due to this development, and collaborative work, the differential relative errors
 in certain chemical abundance measures are now sometimes only a few percent,
 or even less, as has been illustrated in this review. It is important to note then
 that these numbers refer to {\it differential} errors, i.e. abundance differences
 relative to a similar star, like the Sun for solar twins or analogues, or a component in
a binary which is almost identical to its companion. The reason for this
 limitation of the present paper is that hardly any {\it absolute} astrophysical abundance
is known to the prescribed accuracy of better than 0.04\,dex. Even for the Sun,
 the absolute errors, in spite of excellent observed spectra, 3D models and detailed non-LTE
 calculations that can be
checked versus detailed observations of the solar disk, we have still to expect
 somewhat larger absolute errors for almost all elements. This is also underlined
 by an external check that is
available, the comparison with the estimate of the sound speed in the solar interior
 \citep{Basu13,Vinyoles17},  which still deviates significantly from what one would expect
 from the photospheric
abundances of C and O \citep{asplund09, Steffen15, amarsi18b}.

It is also important to realize that accurate absolute abundances are needed in many
 astrophysical applications. In studies of the action of  the s-process
in the interior of AGB stars and the transportation of these elements to the
 surface of the stars, the true amounts of these elements matter for,
 e.g. a proper understanding of their
production in the Galaxy. Similarly, the yields of carbon from AGB stars of
 various types, from Wolf-Rayet stars and from supernovae, and the accurate
 carbon abundances in stars of
different ages and populations in the Galaxy, are needed for a proper
 understanding of the carbon production as a whole. The same holds true for
 many chemical elements, and not the
least regarding their production in the early Galaxy, studies where absolute
 abundances have to be measured for stars very different from the Sun.
 Although relative abundances of for instance
Pop II stars still may reveal interesting structures and give insight into
 the degree of mixing and frequency of supernovae of different types, a
quantitative understanding of such processes will require absolute abundances. Also,
 the study of other processes beyond nucleosynthesis will benefit from absolute abundances of high
quality, and at least free from severe systematic errors.
 Among those processes is ``diffusion'' -- or sooner the interaction of
 radiative pressure, turbulent and other mixing, and sinking in the gravitational field -- 
 in the surface layers
of stars. Also, the consequences of planetary interaction, a subject that has been
 developed efficiently by differential studies as seen above,
 will need absolute abundances when applied
to a wider range of spectral types and to single stars. Of particular interest here
 are the M dwarfs which have so different spectra and atmospheric structures
 that truly differential analyses cannot be extended very far in temperature.

It is certainly easy to ask for more measurements in higher accuracy, 
but is it really worth the effort? A sweeping argument here would be to refer to
William Thomson, 1st Baron Kelvin, who in a famous quote on the need to measure argued
 that what you cannot express in numbers  ``may be the beginning of knowledge, but
you have scarcely ... advanced to the stage of science''
\citep{thomson1889}. A more precise argument
 is that the evolution of the abundance determination for stars have proven to continuously
generate new and unexpected knowledge when the accuracy has been raised.
 This is illustrated above in numerous cases. So, we have no reason to believe that this development
would stop when we take further steps in the development towards high-precision abundances.

Thus, we conclude that the area of absolute abundance determination should be further developed.
  This will require full 3D modelling of stellar atmospheres, and ultimately
also with statistical-equilibrium in the models themselves and not only the spectra,
 and further on with magneto-hydrodynamic effects accounted for properly.
 One important step towards
 such models will also be a detailed and critical comparison with high-resolution
high S/N spectra extending from UV to far IR for a number of different standard stars.

In the continuing efforts to improve the accuracy and advance 
the abundance analysis to new types of stars, the work by atomic and
molecular physicists in improving  the necessary basic data will be of
fundamental significance. Such efforts must be enthusiastically encouraged
and adequately supported by the astrophysics community.

A very impressive development has taken place during recent decades not only
 in the accuracy of the analyses but also in the number of stars analysed.
 In the 1960's, a PhD thesis could contain
the analysis of one single star. Present samples may contain several hundreds,
 and in some cases even thousands of stars, and several tens of elements to be
 analysed for each of them.
Some of these large projects unfortunately suffer from inhomogeneities,
 in instrumentation, reduction methods of the observations, or analysis;
 for a discussion, see for example \citet{jofre17b}.  According to
our experience, it is indeed very important for the usefulness of a study to keep 
the methodology constant. Another important directive when large surveys are to be planned is certainly to
use strict and well defined selection criteria. In order to avoid inhomogeneities
 in methods and samples, merging of different data sets should be avoided (which
 certainly does not mean that one should not inter-compare  the sets critically).

A promising but as yet rather primitive approach towards the very extensive 
and rapidly expanding data sets of stellar spectra, is to develop automatic analysis systems.
 It is our impression that considerable efforts should be made to advance such algorithms 
before they can be left alone, freed from the critical watching eyes of experienced astronomers.
 In particular it must be important to design them to report on stars and stellar groups 
that depart from the typical stars in their training samples.
 The great homogeneous surveys of present-day astronomy
open up the possibilities for making discoveries of new totally unexpected
 phenomena in nature. We should not miss that opportunity.

In planning extensive surveys, it is important to carefully balance the need
 for a large sample of stars against the spectral resolution and the S/N necessary
 for reaching the accuracy aimed for. It 
is not enough to trade a decrease in accuracy in the results for individual stars 
against the formal accuracy in the mean for these stars. Such means may be plagued by 
considerable systematic errors, e.g. in the continuum definition or due to blends
 if the S/N or the resolution are not high enough, see \citet{Lindegren13} for an
 illustrative discussion of this.
In the choice between great samples of stars and relatively resolved spectra
 at high S/N of a smaller number of stars, the latter is often to be preferred
 and can in many cases be realized with small
losses of representativity if care is exercised in the definition of the sample.


\begin{acknowledgements}
The anonymous referee as well as Anish Amarsi, Paula Jofr{\'e}, Karin Lind, and
Jorge Mel{\'e}ndez are thanked for critical reading of a first version of the
manuscript and for many helpful suggestions for improvements. 
Funding for the Stellar Astrophysics Centre is provided by the
Danish National Research Foundation (Grant DNRF106).
\end{acknowledgements}

\bibliographystyle{spbasic}      
\bibliography{Nissen-Gustafsson}   

\end{document}